\documentclass[mathpazo]{cicp}

\usepackage{subcaption}
\usepackage{bm}
\usepackage{algorithm}
\usepackage{algpseudocode}
\usepackage{pythonhighlight}
\usepackage[htt]{hyphenat}
\newcommand{\vecti}[1]{\mathsf{#1}}
\newcommand{\vects}[1]{\MakeUppercase{\vecti{#1}}}
\newcommand{\vect}[1]{\bm{\vecti{#1}}}
\newcommand{\vectn}[1]{N_\vecti{#1}}
\newcommand{\mapnl}[1]{\mathcal{#1}}
\newcommand{\mapl}[1]{\mathsf{#1}}

\begin{document}

\title{COMPUTATIONAL SOFTWARE\vspace{1cm}\\DAFI: An Open-Source Framework for Ensemble-Based \textbf{D}ata \textbf{A}ssimilation and \textbf{F}ield \textbf{I}nversion}

\author[Michelén~Ströfer C A et.~al.]{Carlos A. Michelén~Ströfer, Xin-Lei Zhang, Heng Xiao\corrauth}
\address{Kevin T. Crofton Department of Aerospace and Ocean Engineering, Virginia Tech, Blacksburg, VA 24061, USA}
\emails{{\tt cmich@vt.edu} (C.~Michelén~Ströfer), {\tt hengxiao@vt.edu} (H.~Xiao)}

\begin{abstract}
    In many areas of science and engineering, it is a common task to infer physical fields from sparse observations. 
    This paper presents the DAFI code intended as a flexible framework for two broad classes of such inverse problems: data assimilation and field inversion. 
    DAFI generalizes these diverse problems into a general formulation and solves it with ensemble Kalman filters, a family of ensemble-based, derivative-free, Bayesian methods.
    This Bayesian approach has the added advantage of providing built-in uncertainty quantification. 
    Moreover, the code provides tools for performing common tasks related to random fields, as well as I/O utilities for integration with the open-source finite volume tool OpenFOAM. 
    The code capabilities are showcased through several test cases including state and parameter estimation for the Lorenz dynamic system, field inversion for the diffusion equations, and uncertainty quantification. 
    The object-oriented nature of the code allows for easily interchanging different solution methods and different physics problems. 
    It provides a simple interface for the users to supply their domain-specific physics models. 
    Finally, the code can be used as a test-bed for new ensemble-based data assimilation and field inversion methods. 
\end{abstract}

\ams{35R30, 76M21, 60-04}
\keywords{data assimilation, inverse modeling, random fields, ensemble Kalman filter, Bayesian inference.}

\maketitle

\section*{Program Summary}
\begin{small}
\begin{description}
\item[Program title:] DAFI 
\item[Nature of problem:] This software performs ensemble-based, derivative-free, Bayesian inference of physical fields from sparse observations.
\item[Software licence:] Apache-2.0  
\item[CiCP scientific software URL:]
\item[Programming language(s):] Python
\item[Computer platform:] Any 
\item[Operating system:] Any
\item[Compilers:] 
\item[RAM:] 
\item[External routines/libraries:]
\item[Running time:]
\item[Restrictions:] 
\item[Supplementary material and references:]
\item[Additional Comments:] Runs on any system with Python and NumPy. Running time and computational requirements depend on specific problem being solved.

\end{description}
\end{small}

\section{Introduction}
\label{sec:intro}
    Inverse problems in physical systems take many forms, and two broad classes---data assimilation and field inversion---are considered here.
    \emph{Data assimilation}~\cite{law2015data} refers to a class of inverse problems where a dynamic model is available and time-dependent observations are used to infer some property of a dynamic system. 
    An example of a data assimilation problem is inferring the temperature field of a heated solid at the current time by using both sparse observations of the temperature (i.e. at a few locations) and a model forecast of the entire field. 
    The model forecast could be obtained from propagating the temperature field at an earlier time using the diffusion equation. 
    Here, \emph{Field inversion} problems refer to a class of inverse problems where two sets of fields are related to each other through a forward model and observations of the output fields are used to infer the input fields. 
    With the heat diffusivity example, a field inversion problem is inferring the material diffusivity from steady-state temperature measurements, where the heat diffusion equation is the forward model relating a diffusivity field to a temperature field.
    The approach taken here is to formulate both data assimilation and field inversion problems within a general framework of inverse problems and solve them by using ensemble Kalman filtering methods~\cite{evensen2009data}. 
    This is possible since field inversion problems can be recast as artificial dynamics problem and solved iteratively by using data assimilation procedures~\cite{iglesias_ensemble_2013}. 
    The main intended application for our code, which we named DAFI, is solving field inversion problems described by partial differential equations (PDE), a common type of problem in science and engineering. 
    The code has several features that reflect this emphasis on PDE-based inversion problems, including: (i) ensemble-based solution approaches which are non-intrusive, requiring no gradients from and no code modification to the physics solver, which can therefore be treated as a black-box model, (ii) separation of statistical inference and physical modeling via object-oriented programming that allows the user to focus only on the physics problem at hand, and (iii) modules that facilitate working with discretized random physical fields. 
    
    DAFI is particularly geared to inverse problems involving fields, which are continuous-valued quantities over some domain.
    For example, the authors have used DAFI to infer the Reynolds stress tensor field from sparse observations of the velocity field in the Reynolds-averaged Navier--Stokes equations~\cite{xiao_quantifying_2016,wu2016bayesian}, and to solve for the porosity field in problems involving fluid flow through porous media~\cite{wang2016data}. 
    To facilitate working with fields, DAFI includes modules that perform common field operations, such as calculating the norm of a discretized field, generating covariance matrices, or performing a modal decomposition of a random field. 
    The code also includes a module for reading and writing fields for OpenFOAM, a widely-used open-source finite volume PDE-solver. 
    
    All methods in DAFI are ensemble Kalman filtering approaches, which are particularly suited for large scientific and engineering problems due to their derivative-free and Bayesian nature. 
    A derivative free approach is advantageous, since obtaining the gradient of a cost function with respect to input fields in complex science and engineering models is usually a non-trivial task. 
    It would necessitate either deriving the adjoint equations and implementing an adjoint solver or implementing an intrusive discrete adjoint into the main code. 
    This is even infeasible for commercial solvers where the source code is not accessible. 
    Another benefit of ensemble-based methods is the use of a Monte Carlo representation for all probability distributions. 
    This becomes a necessity for complex fields with large discretization since manipulating and propagating large covariance matrices through physical models quickly becomes intractable. 
    Finally, ensemble Kalman filters are fully Bayesian and result in an estimate of the posterior distribution, which is particularly useful in applications that require quantifying uncertainty in the inferred quantities or fields. 
    
    DAFI  is implemented in an object-oriented manner with every problem requiring two objects: a statistical method for the inference and a physics model describing the particular problem being solved. 
    This separation allows the code to be useful to both physics domain experts trying to solve a particular problem and to algorithm developers allowing them to test their statistical methods. 
    For the physics domain experts DAFI has a simple mechanism to provide the problem-specific physics model. 
    This physics model could be as simple as a non-intrusive wrapper around a third party or commercial solver. 
    Such a user should have domain-specific knowledge but little knowledge is required in the statistical methods used for field inversion. 
    Unlike the models describing the physics, the statistical methods are problem agnostic, and the code provides a library of such methods.
    For the algorithm developer DAFI can act as a test-bed for testing new ensemble-based inversion methods. 
    For example, the authors have used DAFI as an algorithmic test-bed to incorporate soft constraints from prior or physical knowledge through a novel regularized ensemble Kalman filter~\cite{zhang2019regularization}, to enforce boundary conditions during field inversion~\cite{michelenstrofer2019enforcing}, and to enforce additional PDEs through physics-informed covariance~\cite{wu2019physics}. 

    The remainder of this introduction provides more details on data assimilation and field inversion, introduces uncertainty quantification problems, and provides a review of other existing codes for data assimilation. 
    Section~\ref{sec:general} presents the general problem formulation and shows how each of the problems presented here can be formulated in this framework.
    In Section~\ref{sec:methods} the statistical methods used to solve the general problem are presented.
    The code implementation is described in Section~\ref{sec:implementation}.
    The treatment of random fields in the DAFI code is described in Section~\ref{sec:fields}. 
    Section~\ref{sec:cases} presents several test cases to illustrate the use of the DAFI code.
    Finally, Section~\ref{sec:conclusion} concludes the paper.

    \subsection{Data Assimilation in Dynamic Systems}
    \label{sec:intro:da}
        In DAFI, field inversion problems are solved by using data-assimilation procedures iteratively and data assimilation techniques are therefore important for solving both data assimilation and field inversion problems. This subsection introduces data assimilation problems and relevant nomenclature.
        Fundamentally, data assimilation consists of inferring the state and/or parameters of a dynamic system by using a dynamic model and observations.
        Using the heat diffusion example, the observations can be spatially sparse temperature measurements (i.e. only at a few locations) at different times while the dynamic model is the diffusion equation. 
        The observations and model can be used to infer the state (e.g. temperature) at a given time, the initial or boundary conditions, or some constant model parameter (e.g. diffusivity). 
        Data assimilation problems are usually classified as either state estimation or parameter estimation, where the state is dynamically varying (e.g. temperature) and parameters are for instance model constants or material properties (e.g. diffusivity). 
        However, the same solution techniques work for both state estimation and parameter estimation problems. 
        In the case where both the state and parameters are being inferred, the approach is to combine both into an \emph{augmented state}~\cite{evensen2009ensemble}.
        Hereafter, \emph{state} shall refer to all quantities being inferred, which may include the true state, model parameters, or both. 
        In problems involving fields, the continuous fields must be discretized to be included in the state vector to be inferred. 
        Alternatively, reduced order modeling, e.g. based on modal decomposition, can be used to represent the field with a finite set of mode coefficients. 
        Section~\ref{sec:fields} describes the treatment of fields in the DAFI code.
        
        \begin{figure}[!htb]
            \centering
            \includegraphics[scale=0.54]{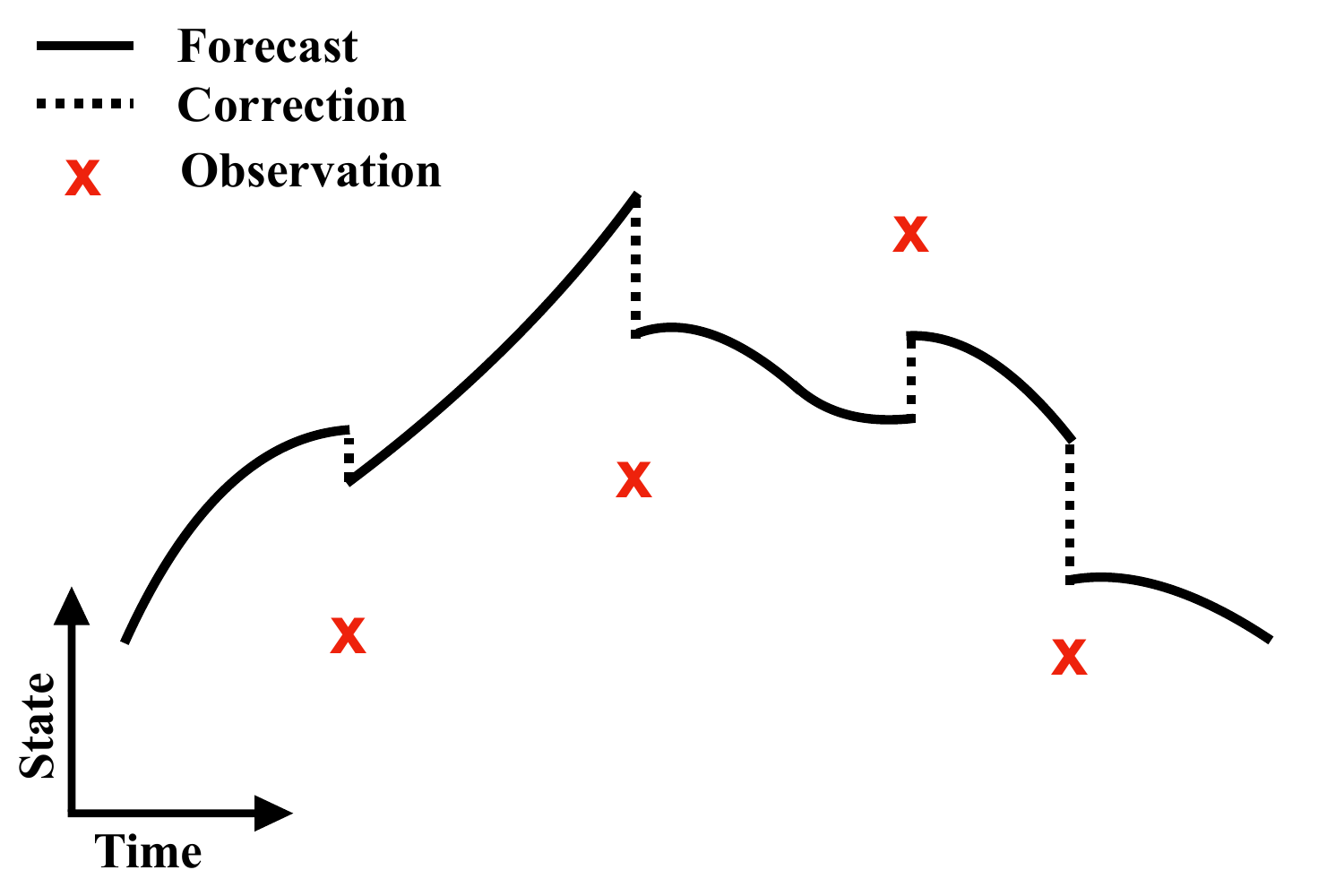}
            \caption{Illustration of a simple filtering data assimilation problem consisting of a single scalar state. The dynamic model is used to forecast the state to the next observation time. At each observation time a linear filter is used to correct the forecast state to the analysis state using the observations. In this illustrative problem, the state space and observation space are the same which is not generally the case for realistic problems.}
            \label{fig:filter}
        \end{figure}
        
        The state estimation problem can be solved in different ways, two of which are filtering and smoothing. 
        In the \emph{filtering} approach the state vector is corrected each time an observation is made by using the observations and the model \emph{forecast}.
        Starting from the initial condition, the dynamic model is used to propagate the state to the first time at which there are observations.
        The observations are then used to correct the forecast state to an \emph{analysis} state using a linear filter.
        Starting with the analysis state as the new initial condition, the dynamic model is used to propagate the state to the next time at which there are observations.
        This is illustrated in Figure~\ref{fig:filter}. 
        The \emph{smoothing} approach is characterized by the use of data at future times to infer an earlier state. 
        There are different smoothing approaches and the smoothing problem can be formulated as a sequential method similar to the filtering problem~\cite{evensen2000ensemble}. 
        Both types of problems are common in data assimilation and in principle the DAFI code is well suited for solving smoothing problems but the code has not yet been used for such problems. 
        The rest of this paper focuses on the filtering problem, whose formulation is also used to solve the field inversion problem iteratively.

    \subsection{Field Inversion}
    \label{sec:intro:fi}
        Field inversion problems consist of inferring some input fields from (possibly sparse) observations of output fields, where the input and output fields are related through a non-linear \emph{forward} model. 
        Like the dynamic model earlier, the forward model generally consists of a system of coupled PDEs. 
        Field inversion problems can be solved using data assimilation techniques iteratively. The iterations are required to account for the non-linearity of the forward model since data assimilation filtering techniques assume a linear mapping between the state and observations. 
        This is presented in more detail in Section~\ref{sec:methods:enkf:ienkf}. 
        
        The main advantage of inferring the input fields over inferring the output fields directly is that the physics described by the forward model is enforced on the output field. 
        Additionally, often times the forward model captures the correct physics but there are uncertainties in some input fields. 
        As an example, Xiao et al.~\cite{xiao_quantifying_2016} infer the Reynolds stress field from sparse observations of the velocity field in computational fluid dynamics simulations.
        The forward model consists of the Reynolds-averaged Navier--Stokes (RANS) equations which propagate a Reynolds stress field to a velocity field. 
        While the RANS simulations capture the correct physics, the Reynolds stress field requires modeling and is the largest source of uncertainty in RANS simulations.
        Inferring Reynolds stress tackles the real source of uncertainty and ensures the output fields (velocity, pressure) satisfy the physics described by the RANS equations.

    \subsection{Uncertainty Quantification}
    \label{sec:intro:uq}
        In some cases simply updating the state or parameters based on observations is not enough, and an estimate of the confidence in these new estimates is also sought. 
        The methods implemented in DAFI are ensemble-based Bayesian approaches and hence always result in a full posterior distribution.
        Therefore, in DAFI, all data assimilation and field inversion problems are also uncertainty quantification problems. 
        For uncertainty quantification, one could for example use the mean of the posterior distribution as the updated state and the covariance as a measure of its uncertainty. 
        
        Bayesian updating provides a posterior probability distribution from a combination of a prior distribution and a likelihood derived from the observations. 
        As an example, if the diffusivity parameter in the heat diffusion problem is being inferred, we could specify the prior distribution---our belief and confidence on the value of diffusivity prior to seeing the data---based on the mean and standard deviation of the material (e.g. steel).
        The measured values (e.g. point-measurements of temperature) and instrumentation's uncertainty determine a likelihood distribution for the state, that is, the probability that the observed measurements would result from a given diffusivity parameter. 
        The Bayesian update then produces a posterior distribution giving the probability of any value of diffusivity given the observations. 
        The mean of the posterior distribution can be used as the updated diffusivity and the variance as a measure of its uncertainty.

    \subsection{Review of Existing Data Assimilation Codes}
    \label{sec:intro:other}
        A number of free, open-source codes have been developed for data assimilation and are reviewed here. 
        The two common approaches to solving data assimilation problems are ensemble data assimiliation, based on the ensemble Kalman filter, and variational data assimilation, a gradient-based approach.
        The National Center for Atmospheric Research (NCAR) created the DART~\cite{anderson2009data} code as a community testbed to try out different data assimilation methods. 
        This platform is the most developed and widely used platform for ensemble data assimilation. 
        It is used both in research and in deployed operational cases and includes good parallelisation options.
        The code is frequently updated to include the latest methods in the literature. 
        The main downside of DART is the steep learning required from the user. 
        In many science and engineering applications a light-weight code would be more accessible to physics domain experts who are not necessarily experts in data assimilation or statistical inference. 
        
        OpenDA~\cite{verlaan2010openda} is another open-source toolkit for data assimilation. 
        It has both ensemble and variational data assimilation methods, as well as methods for uncertainty analysis and model calibration. 
        One of the driving philosophies of the project is to be a platform for researchers to share and exchange code and ideas.
        As such, the code is object-oriented and users can share both new models and new data-assimilation techniques for use by the wider community. 
        OpenDA has found many applications in ocean engineering, with the code coupled to popular codes such as OpenFOAM and SWAN. 
        OpenDA is similar to DART in that it is very general, powerful, and well-suited for large deployment scale problems. 
        As a result, it shares the same drawback as DART, the steep learning cost associated with it. 
        For instance, OpenDA has language interfaces with C/C++, Java, and Fortran. 
        While these are good options for large scale deployment, usually engineering research and exploration happens in more user-friendly interpreted languages such as Python or Matlab. 
        The use of compiled languages can be a major overhead to many researchers. 
        
        DATeS~\cite{attia2017dates} is another code developed at Virginia Tech as a testbed for new data assimilation algorithms. 
        Its implementation shares many similar philosophies with DAFI, including using Python object-oriented programming to easily switch between models or data assimilation techniques. 
        DATeS is capable of using both ensemble and variational methods. 
        In addition to different data assimilation methods, DATeS has the ability to switch between different linear algebra and time integration routines. 
        Many of these features are incorporated because of the focus on testing and evaluating the performance of different algorithms. 
        DATeS is more accessible to a naive user than DART but still includes some complexity due to its generality and focus on algorithm testing. 
        
        The main distinctive attributes of our code are its specialization to ensemble methods, its ease of use, and its field operations. 
        These attributes reflect the focus on engineering and physics applications and make DAFI an accessible code for solving diverse problems by physics domain experts who are less versed in statistical inference. 
        Specific examples of this focus include the modal decomposition of random fields for reduced order modeling and the ability to account for unstructured meshes when performing field calculations. 
        These features of the code were chosen based on the authors' experience with field inversion in diverse problems, including turbulence modeling, tsunami-induced sediment transport, vegetation modeling in coastal engineering, and flow through porous media. 
        Admittedly, the other codes reviewed here are capable of using more general solution approaches than DAFI, but this generality, however, comes at the cost of simplicity and ease of use. 
        For instance, by specialising in ensemble methods, DAFI always requires the same small set of functions from any physics model. 
        This requires little understanding of data assimilation techniques from the part of the user, and makes it simple to couple their physics model. 
        Similarly, because of the specialization to ensemble methods, adding new ensemble methods is straightforward. 
        Ensemble methods are generalized in DAFI to consists of two loops, a time marching outer loop and an inner loop that iterates at the same time, a structure that can encompass all ensemble methods the authors are aware of. 
        This structure and functionalities such as checking convergence and saving intermediate results are already implemented in a general class and adding a new method requires only overriding the update scheme.

\section{General Problem Formulation}
\label{sec:general}
\begin{figure}[!htb]
        \centering
        \includegraphics{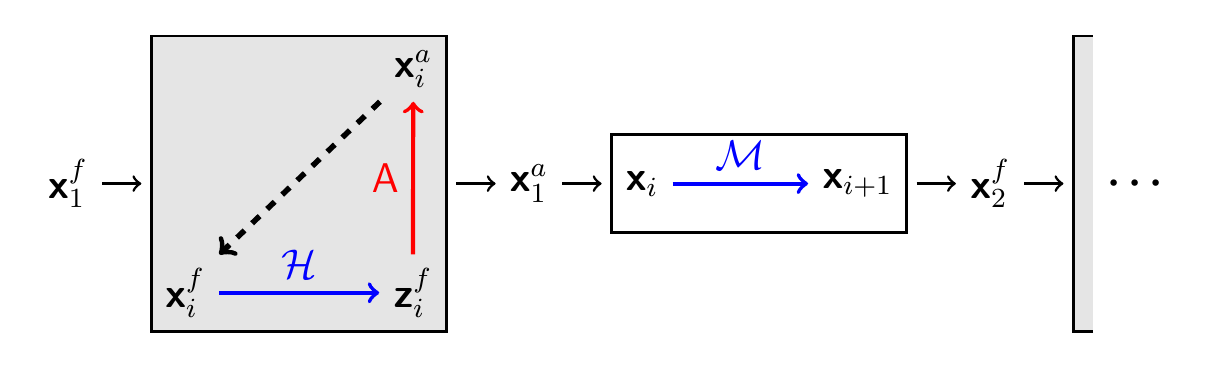}
        \caption{Schematic of the general problem formulation. Starting with the forecast state at the first data observation time $\vect{x}^f_1$, the state is first updated via data assimilation (grey box) to the analysis state at the same time $\vect{x}^a_1$. The analysis state is then forecast to the next observation time $\vect{x}^f_2$ via the dynamic model $\mapnl{M}$ (white box). This sequential data assimilation and forecasting process continues for each observation time. In the data assimilation step (grey box), the forecast state $\vect{x}^f$ is updated to the analysis state $\vect{x}^a$ through the iterative update scheme that includes the nonlinear observation operator $\mapnl{H}$ and the linear filter $\mapl{A}$. The linear filter $\mapl{A}$ takes three inputs: the forecast state $\vect{x}^f$, the same state mapped to observation space $\vect{z}^f$, and the observations at the current data assimilation time $y$ (not shown). The dynamic model and observation operators shown in blue are implemented in the user-defined physics model while the filter shown in red is chosen as one from the library contained in the DAFI \texttt{inverse} module. The same color scheme is used in Algorithm~\ref{alg:DA}.} 
        \label{fig:filter2}
    \end{figure}

    A general problem is formulated here to encompass all the different classes of problems described above.
    The problem is implemented as a filtering approach but with two distinguishing features: (1) the observation operator is separated from the filter and allowed to be non-linear, and (2) the analysis step is done iteratively.
    This means the solver consists of two loops: an outer time-marching loop and an inner iteration loop. 
    The inner loop is used to deal with non-linearity either in the observation operator (e.g. field inversion problem), in the dynamic model (e.g. the ensemble randomized maximum likelihood method for highly nonlinear models), or both. 
    The process is summarized in Figure~\ref{fig:filter2} and Algorithm~\ref{alg:DA}, which are further described below. 
    The details of the ensemble-based Bayesian solution approach for this general problem are deferred until Section~\ref{sec:methods}. 

    A system consists of a state vector $\vect{x}\in\vects{x}$ with dimension $\vectn{x}$ whose evolution in time $t$ is described by a nonlinear dynamic model $\mapnl{M}:\vects{x}\mapsto\vects{x}$.
    The state vector refers to the vector of variables being inferred, which can generally be an augmented state vector.
    Observations of the system consist of pairs $\left\{(t_i, \vect{y}_i)\right\}_{i=1}^{N_t}$, where $t_i$ correspond to observation times and $\vect{y}_i\in\vects{y}_i$ is the observation vector with dimensions ${\vectn{y}}_i$.
    In general the observation space is different at each observation time since different quantities might be observed at each time.
    The state space is mapped to the observation space through a nonlinear observation operator $\mapnl{H}_i:\vects{x}\mapsto\vects{y}_i$.
    For simplicity of notation the subscript $i$ will be dropped except where the distinction is necessary, but it should be noted that the observation spaces at different times are generally distinct and hence so are the operators $\mapnl{H}$ and $\mapl{A}$.
    The state mapped to observation space is given by
    \begin{equation}
        \vect{z}=\mapnl{H}(\vect{x}) \in \vects{y} \text{.}
        \label{eq:Hx}
    \end{equation}
    The analysis step is given by
    \begin{equation}
        \vect{x}^\text{a} = \mapl{A}(\vect{x}^\text{f}, \vect{z}^\text{f}, \vect{y}) \text{,}
    \end{equation}
    where $\vect{x}^\text{f}$ and $\vect{x}^\text{a}$ are the forecast and analysis states, and $\mapl{A}:(\vects{x}\times\vects{y}\times\vects{y})\mapsto\vects{x}$ is the linear filter.
    An iteration is added where the analysis state becomes the forecast step.
    This is indicated by the dashed lines in Figure~\ref{fig:filter2}, and the loop is repeated until some convergence criteria are met.
    Different filtering techniques and how they fit within this general formulation will be discussed in Section~\ref{sec:methods}.
    After the filtering loop is complete, the \emph{dynamic model} is used to propagate the state to the next observation time as
    \begin{equation}
        \vect{x}_{i+1} = \mapnl{M}(\vect{x}_i) \text{.}
        \label{eq:dynamic}
    \end{equation}
    Here, $\vect{x}_i$ indicates the state at the time corresponding to the $i$\textsuperscript{th} observation and the dynamic model has its own time-stepping scheme to take the state from $\vect{x}_i$ to $\vect{x}_{i+1}$. 
    This procedure is summarized in Algorithm~\ref{alg:DA}.

    \begin{algorithm}
      \caption{General Problem Formulation}
      \label{alg:DA}
      \begin{algorithmic}[1]
        \Procedure{DataAssimilation}{$\vect{x}_{1}^\text{f}$, $\{(t_i,\vect{y}_i)\}_{i=1}^{N_t}$} \Comment{outer loop}
            \For{$i = 1$ \textbf{to} $N_t$}
                \State $\vect{x}_{i}^\text{a} \gets \operatorname{Analysis}(\vect{x}_i^\text{f}, \vect{y}_i)$ \Comment{analysis state}
                \State $\vect{x}_{i+1}^\text{f} \gets \textcolor{blue}{\operatorname{DynamicModel}}\left(\vect{x}_i^\text{a}, t_i, t_{i+1} \right)$ \Comment{forecast state}
            \EndFor
        \EndProcedure
        \State
        \Procedure{Analysis}{$\vect{x}_i, \vect{y}_i$} \Comment{inner loop}
            \While{\texttt{not converged}}
                \State $\vect{z}_i \gets \textcolor{blue}{\operatorname{ObservationOperator}}(\vect{x}_{i})$
                \State $\vect{x}_i \gets \textcolor{red}{\operatorname{Filter}}\left( \vect{x}_i, \vect{z}_i, \vect{y}_i \right)$
            \EndWhile
            \State \textbf{return} $\vect{x}_i$
        \EndProcedure
      \end{algorithmic}
    \end{algorithm}

    The different problems described in Section~\ref{sec:intro} can be described using this general formulation. 
    These are summarized in Table~\ref{tab:cases} and described in the following list:
    \begin{description}
        \item[Filtering] The inner loop is not used and the outer loop is advanced by the dynamic model. 
        \item[Field Inversion] The outer loop is not used and the inner loop iterations are used to account for the non-linearity of the problem.
        In this case the state to be inferred is the input field, and the observation operator consists of two components: the forward model that maps from input field (state) to output field (observable state) and an observation operator on the observable state. 
        For observable state $\vect{u}\in\vects{u}$, forward model $\mapnl{F}:\vects{x}\mapsto\vects{u}$, and observation operator on the observable state $\mapnl{H}_{\vect{u}}$, the observation operator is given by $\mapnl{H}(\vect{x}) = \mapnl{H}_{\vect{u}}\left(\mapnl{F}(\vect{x})\right) = \mapnl{H}_{\vect{u}}\left(\vect{u}\right)$.
    \end{description}

    While it is useful to consider each type of problem separately, they all fit into the general formulation where the outer loop deals with time marching and data-assimilation steps, the inner loop is used to address any non-linearity, and the general observation operator maps from state space to observation space. 
    In previous works, the authors have used the DAFI code for classical filtering, field inversion, and uncertainty quantification problems, and these are showcased in Section~\ref{sec:cases}. 

    \begin{table}[ht]
        \centering
        \caption{Implementation of specific problems using the general formulation. For each type of problem it is indicated whether the outer and inner loops are iterated or passed through only once. The form of the observation operator is also shown.}
        \label{tab:cases}
        \begin{tabular}{c|c|c|c}
            Problem & Outer Loop ($\mapnl{M}$) & Inner Loop ($\mapl{A}$) & Observation Operator \\
            \hline
            Filtering & iterate & once & $\mapl{H}$ \\ 
            Field inversion & once & iterate & $\mapnl{H}_{\vect{u}}\circ\mapnl{F}$ \\ 
            General & iterate & iterate & $\mapnl{H}$ 
        \end{tabular}
    \end{table}

\section{Ensemble-Based Analysis Methods}
\label{sec:methods}
    In the general problem above, the analysis step consists of combining the model prediction (forecast state) and the observations to obtain an improved prediction (analysis state).
    The approach taken is a Bayesian implementation, where both the model prediction and the observation are considered random vectors.
    The state vector is now a random vector possibly consisting of discretized random fields.
    The goal is then to obtain the probability distribution of the analysis state.
    In the language of Bayesian inference, the forecast state is the \emph{prior} probability distribution $P(\vect{x})$, the observations imply a \emph{likelihood} distribution $P(\vect{y} \mid \vect{x})$ for any proposed state vector, and the analysis state is the \emph{posterior} probability distribution $P(\vect{x} \mid \vect{y})$.
    The resulting posterior distribution is given by Bayes' formula
    \begin{equation}
        P(\vect{x} \mid \vect{y}) \propto P(\vect{x}) P(\vect{y} \mid \vect{x}) \label{eq:bayes} \text{,}
    \end{equation}
    where the constant of proportionality ensures a total probability of one, i.e. $\int_{\vect{x}}P(\vect{x} \mid \vect{y})=1$. 
    The derivation of the ensemble Kalman filter assumes a Gaussian distribution for all random vectors (Gaussian process for random fields), even though admittedly the distributions do not remain Gaussian after propagation through a non-linear dynamic model. 

    Random vectors with Gaussian distributions are completely characterized by a mean vector and a covariance matrix.
    For our general problem, the Bayesian formulation requires the forecast state to consist of a mean value and a covariance matrix. 
    For steady-state problems this means defining a prior distribution, that is a prior mean and prior covariance matrix.
    For dynamic systems this can be obtained by defining the initial condition as a Gaussian distribution, and propagating this distribution to the first data assimilation time using the dynamic model.
    The analysis step then modifies this propagated (forecast) distribution using Bayes' theorem.
    This updated (analysis) distribution is then propagated again to the next data assimilation time, and so on. 
    The observations at each time are also considered Gaussian distributions, with mean equal to the measurement values and variances obtained from the accuracy of the measuring instruments.
    Observations are typically considered independent of each other resulting in a diagonal covariance matrix.

    With a Guassian assumption and linear observation operator $\mapl{H}$, Bayes' formula results in a Gaussian posterior distribution given by the following mean and covariance~\cite{welch1995introduction}
    \begin{subequations}
        \begin{align}
            & \overline{\vect{x}}^a = \overline{\vect{x}}^f + \mapl{K}(\overline{\vect{y}}-\mapl{H}\overline{\vect{x}}^f)  \text{,} \\
            & C_{\vect{x}}^a =  (\mathrm{I}-\mapl{K}\mapl{H})C_{\vect{x}}^f \text{,}
        \end{align}
        \label{eq:kf1}
    \end{subequations}
    where the posterior is relabeled as the analysis state $\vect{x}^a=\vect{x} \mid \vect{y}$ and the prior is relabeled as the forecast state $\vect{x}^f=\vect{x}$.
    Overlines denote the mean value of the distribution, $C_{\vect{x}}$ denotes the covariance matrix of the state vector, and $\mapl{K}$ is the Kalman gain matrix given by
    \begin{equation}
        \mapl{K} = C_{\vect{x}}^f\mapl{H}^\top\left(\mapl{H}C_{\vect{x}}^f\mapl{H}^\top + C_{\vect{y}} \right)^{-1} \text{,}
        \label{eq:kf2}
    \end{equation}
    where $C_{\vect{y}}$ is the covariance matrix of the observations.
    This procedure is the Kalman filter~\cite{welch1995introduction}, a common data assimilation technique.
    The Kalman filter is also the basis for more complex methods, e.g., the extended Kalman filter (EKF)~\cite{jazwinski2007stochastic} where the mean state is propagated with the full nonlinear model instead of with the tangent linear model, and the unscented Kalman filter (UKF)~\cite{wan2001unscented} where the mean state and covariance are estimated from a set of propagated samples which are selected based on the unscented transform.
    In the Kalman filter, the posterior mean takes into account not only the mean of the prior and observations but also the confidence in each as defined by their covariance.

\begin{figure}[!htb]
        \centering
        \includegraphics[scale=0.5]{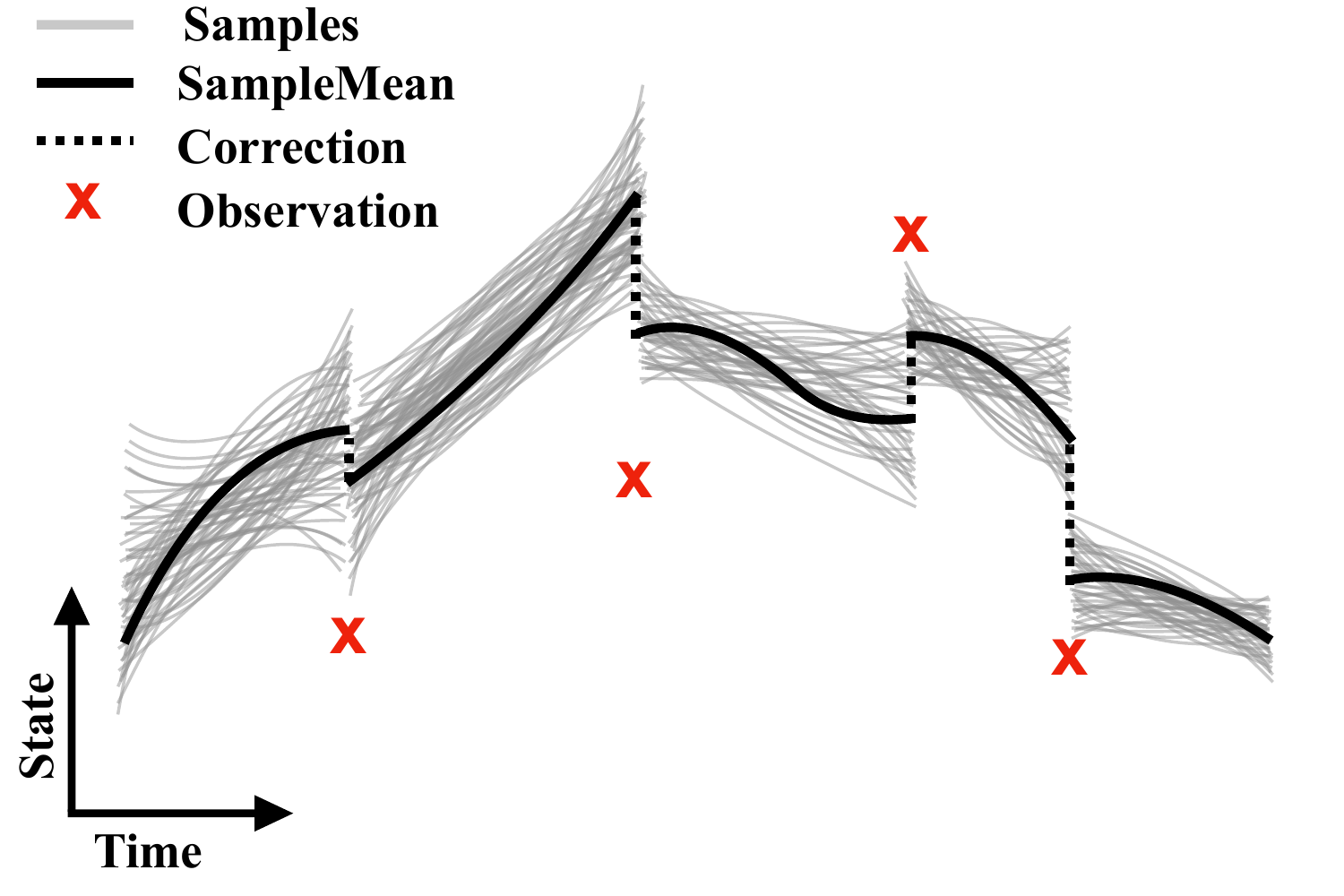}
        \caption{Example of a simple ensemble-based data assimilation problem consisting of a single scalar state. The procedure is similar to that in Figure~\ref{fig:filter} but with an ensemble of states being propagated and updated. The ensemble is a Monte Carlo representation of the probability distribution and the ensemble statistics are used for the data assimilation update. During the data assimilation each sample is updated.}
        \label{fig:ensemble}
    \end{figure}

    Problems involving fields have very large state vectors, proportional to the discretization size.
    Because of this, the state covariance matrix becomes computationally unmanageable and a direct application of the Kalman filter is unfeasible.
    Ensemble data assimilation methods are Monte Carlo approaches where all probability distributions are represented with a finite set of samples.
    The modified distributions (e.g. propagated in time, or after the analysis step) are then described by the sample mean and sample covariance of the modified samples.
    This procedure is illustrated in Figure~\ref{fig:ensemble}.
    A direct application of the Kalman filter using an ensemble results in the ensemble Kalman filter (EnKF)~\cite{evensen2009ensemble}, a common ensemble method and the basis for more complex ensemble methods \cite{gu2007iterative,emerick2013ensemble,bocquet2014iterative}.
    In ensemble methods it is common to perturb the observations for each sample as well~\cite{burgers1998analysis}. 
    For the $j$\textsuperscript{th} sample this is given as 
    \begin{equation}
        \vect{y}^{(j)} \sim \mathcal{N}(\vect{y},C_{\vect{y}}) \text{.}
    \end{equation}
    DAFI has a library of different ensemble-based Bayesian methods.
    The subsections in this section present a brief theoretical background for the different methods which are currently implemented.

    Field inversion problems which involve a non-linear operator can be solved by iterative use of the EnKF. 
    Even for data assimilation problems, if the dynamic model is highly non-linear, each data assimilation step can be solved iteratively. 
    Iterative methods require a convergence or stopping criteria and the correct choice of criteria can affect the method's performance~\cite{kaltenbacher2008iterative,iglesias_ensemble_2013}.
    Currently, DAFI implements two convergence criteria in addition to a specified maximum number of iterations.
    Both convergence criteria are based on the norm of the average (over all samples) misfit, which at iteration $l$ is given as 
    \begin{equation}
        g_{(l)} = \left\lVert \overline{\mapnl{H}(x) - y}\right\rVert_{(l)} \text{.}
    \end{equation}
    The first criteria is based on the iterative residual of the norm of the misfit.
    At the first iteration that this value decreases by less than a specified relative amount $\varepsilon$ the iterations are stopped, that is, iterations are stopped when
    \begin{equation}
      g_{(l-1)} - g_{(l)} \leq \varepsilon g_{(0)} \text{,}
    \end{equation}
    The second stopping criteria is based on the discrepancy principle~\cite{kaltenbacher2008iterative, schillings2018convergence, iglesias_ensemble_2013}.
    The idea behind the discrepancy principle is that once the noise for the misfit is within the same order of magnitude as the noise level of the observation data, the data provides no more information. 
    Continuing with data assimilation could lead to fitting the noise of the data (over-fitting). 
    The discrepancy principle implemented in DAFI is based on the form used by Schillings and Stuart~\cite{schillings2018convergence}, and consists of stopping the iterations when
    \begin{equation}
      g_{(l)} \leq \tau \sqrt{\operatorname{trace}(C_{\vect{y}})} \text{,}
    \end{equation}
    for a $\tau\geq1$. 
    Here the square root of the trace of the observation covariance is the expected value of the norm of the observation noise~\cite{schillings2018convergence}.

    \subsection{Ensemble Kalman Filter (EnKF)}
    \label{sec:methods:enkf}
        The ensemble Kalman filter (EnKF) is a Monte Carlo implementation of the Kalman filter described above~\cite{welch1995introduction}.
        For the filtering problem in Figure~\ref{fig:filter}, the initial forecast state is replaced with an ensemble of $N_s$ samples of the state vectors.
        This ensemble is created, for instance, by choosing a mean and covariance for the initial conditions at $t_0$ and sampling the initial condition using a Gaussian distribution.
        Each sample is then propagated with the dynamic model to the first data assimilation time $t_1$.
        The analysis step then consists of updating each sample individually as
        \begin{subequations}
            \begin{gather}
                \vect{x}^{a(j)} = \vect{x}^{f(j)} + \mapl{K}\left(\vect{y}^{(j)}-\vect{z}^{f(j)}\right) \label{eq:enkf1:1} \text{,} \\
                \vect{z} = \mapl{H}\vect{x} \text{,}
            \end{gather}
            \label{eq:enkf1}
        \end{subequations}
        where the Kalman gain is as in Equation~\eqref{eq:kf2} but with the sample covariance matrix as an estimate for the true covariance matrix.
        Using the definition of sample covariance, the Kalman gain matrix can be written as (see Appendix~\ref{app:kalmangain})
        \begin{equation}
            \mapl{K} = C_{\vect{x}\vect{z}} \left(C_{\vect{z}} + C_{\vect{y}} \right)^{-1} \text{,}
            \label{eq:enkf2}
        \end{equation}
        where $C_{\vect{x}\vect{z}}$ is the covariance matrix between vectors $\vect{x}$ and $\vect{z}$.
        This avoids constructing the large covariance matrix $C_{\vect{x}}$ for the state vector.
        This formulation also avoids explicitly constructing the matrix $\mapl{H}$ by using the vector $\vect{z}=\mapl{H}\vect{x}$ directly.
        Each updated sample $\vect{x}_j^a$ is then propagated with the dynamic model to the next data assimilation time step ($t_2$) and so on.
        As presented here, the EnKF solves the filtering data assimilation problem.
        The next subsection shows how to use the EnKF for field inversion problems.

        \subsubsection{Iterative EnKF}
        \label{sec:methods:enkf:ienkf}
            Problems involving a nonlinear observation operator can be recast as a linear problem using an augmented state and artificial dynamics.
            Iglesias et al.~\cite{iglesias_ensemble_2013} used this approach for field inversion problems by recasting the problem as an artificial dynamics problem.
            The augmented state vector $\widetilde{\vect{x}}$ consists of the state and the state mapped to observation space
            \begin{equation}
                \widetilde{\vect{x}} = \begin{bmatrix} \vect{x}\\ \mapnl{H}(\vect{x}) \end{bmatrix} = \begin{bmatrix} \vect{x}\\ \vect{z} \end{bmatrix} \text{.}
            \end{equation}
            The artificial dynamics model is given by
            \begin{equation}
                \widetilde{\vect{x}}_{i+1} = \widetilde{\mapnl{M}}(\widetilde{\vect{x}}_i) = \begin{bmatrix} \vect{x}_i \\ \mapnl{H}_i(\vect{x}_i) \end{bmatrix} \text{.}
            \end{equation}
            The observation operator is given by $\mapl{H}=\begin{bmatrix}\bm{0} & \mathrm{I}\end{bmatrix}$.
            The problem is then solved iteratively as shown in Figure~\ref{fig:filter2} using the EnKF as in Equations~\eqref{eq:enkf1}-\eqref{eq:enkf2} with the iterations acting as pseudo-time.
            The problem has been recast as an artificial dynamic problem with non-linear dynamic model and linear observation operator which allows for the direct use of the Kalman filter.
            However, with these definitions for the dynamic model, augmented state vector, and observation operator, the resulting Kalman update can be rewritten in terms of the original state vector (see Appendix~\ref{app:iterativekalman}).
            The resulting equations are exactly Equation~\eqref{eq:enkf1:1} and Equation~\eqref{eq:enkf2} but with
            \begin{equation}
                \vect{z}=\mapnl{H}(\vect{x}) \text{.}
            \end{equation}
            The iterative EnKF is not implemented as a separate method from the EnKF, rather the user simply specifies a number of inner loop iterations or convergence criteria. 
            The standard EnKF corresponds to a single inner loop iteration. 
            For all methods the user-provided physics model is queried for the values of $\vect{z}$ 
            which can be obtained by first running a non-linear forward model. 
            When used for non-linear inversion this iterative EnKF approach is sometimes referred to as ensemble Kalman inversion (EKI).

    \subsection{Regularized EnKF (REnKF)}
    \label{sec:methods:renkf}
        Inverse problems are often ill-posed with many different states resulting in satisfactory agreement with the observations.
        This necessitates the introduction of regularization techniques into ensemble methods to further constrain the inference process.
        This is a common technique in variational approaches, but one the traditional EnKF lacks.
        Zhang et al.~\cite{zhang2019regularization} address this complication by augmenting the EnKF to allow general regularization terms in its implied cost function, resulting in the regularized EnKF (REnKF).
        This method is capable of regularizing the inference with additional constraints to enforce a-priori knowledge or preferred behavior.

        A general constraint function on the state $\vect{x}$ enforcing some desired property can be expressed as
        \begin{equation}
          \left\lVert\mapnl{G}(\vect{x})\right\rVert_{W}=0 \text{,}
        \end{equation}
        where $\mapnl{G}$ is a constraint function and $W$ is weight matrix defining the norm to minimize.
        The update scheme of the regularized EnKF is given as
        \begin{subequations}
            \begin{gather}
                \hat{\vect{x}}^{f(j)} = \vect{x}^{f(j)} - C_{\vect{x}}\mapnl{G}^\prime(\vect{x}^{f(j)})^\top \mapl{W} \mapnl{G}(\vect{x}^{f(j)}) \label{eq:renkf1:0} \text{,} \\
                \hat{\vect{z}}^{f(j)} = \mapl{H}\hat{\vect{x}}^{f(j)} \text{,} \\
                \vect{x}^{a(j)} = \hat{\vect{x}}^{f(j)} +  \mapl{K}\left(\vect{y}^{(j)}-\hat{\vect{z}}^{f(j)}\right) \label{eq:renkf1:1} \text{,}
            \end{gather}
            \label{eq:renkf1}
        \end{subequations}
        where $\mapnl{G}^\prime(\vect{x})$ is the derivative of the constraint function $\mapnl{G}$ with respect to the state $\vect{x}$.
        Equation~\eqref{eq:renkf1:0} represents a correction based on enforcing the constraint while Equation~\eqref{eq:renkf1:1} is the standard Kalman update correction based on matching the observations.
        The regularized EnKF can also be used for field inversion problems in the same manner as the EnKF, by recasting the problem as an artificial dynamics and solving iteratively.

    \subsection{EnKF-MDA}
    \label{sec:methods:mda}
        In certain scenarios EnKF leads to overcorrection on the state variables in early iterations.
        This is the case for instance in schemes where the dynamic model (or artificial dynamics) is highly nonlinear and the prior is far from the truth.
        This is due to the EnKF performing the Gauss--Newton update with a full step where the system model is linearized and the averaged sensitivity matrix is used~\cite{reynolds2006iterative}.
        To address this issue, a common approach is to damp the update at early iterations.
        Motivated by this, the ensemble Kalman filter with multiple data assimilation (EnKF-MDA)~\cite{emerick2013ensemble} was proposed to reduce the update amount in each iteration by inflating the observation error covariance. 
        EnKF-MDA performs a single Kalman update but in multiple steps where the same data is used in each step. 
        This corresponds to re-expressing Bayes' formula in Equation~\eqref{eq:bayes} as 
        \begin{equation}
            P(\vect{x} \mid \vect{y}) \propto P(\vect{x}) \prod_{l=1}^{N_\text{mda}} P(\vect{y} \mid \vect{x}_l)^{\frac{1}{\alpha_l}} \text{.}
        \end{equation}
        The EnkF-MDA has been shown to outperform EnKF for nonlinear applications~\cite{emerick2012history,evensen2018}.
        
        The EnKF-MDA is an iterative method with update scheme given as 
        \begin{subequations}
            \begin{gather}
                \vect{x}_{l+1}^{(j)} = \vect{x}_{l}^{(j)} + \hat{\mapl{K}} \left(\hat{\vect{y}}^{(j)}-\vect{z}_{l}^{(j)}\right) \label{eq:enkfmda1:1} \text{,} \\
                \hat{\vect{y}} = \vect{y} + \sqrt{\alpha_l}\epsilon \text{,} \\
                \hat{\mapl{K}} = C_{\vect{x}\vect{z}} \left(C_{\vect{z}} + \alpha_l C_{\vect{y}} \right)^{-1} \text{,} \\
                \vect{z}_{l}^{(j)} = \mapl{H}\vect{x}_{l}^{(j)} \text{,}
            \end{gather}
            \label{eq:enkfmda1}
        \end{subequations}
        where $l$ denotes the sub-iteration index in one data assimilation window, $\alpha_l$ is the inflation parameter at the $l$\textsuperscript{th} iteration step, $\epsilon \sim \mathcal{N}(0, C_{\vect{y}})$ is the observation noise, and $\hat{\mapl{K}}$ is the Kalman gain matrix with inflated observation error.
        The inflation parameters are chosen such that 
        \begin{equation}
          \sum_l^{N_\text{mda}}\frac{1}{\alpha_l}=1 \text{,} \label{eq:enkfmda:inflation}
        \end{equation}
        where $N_\text{mda}$ is a specified number of data assimilation iterations.
        One option is to make the inflation parameters constant, i.e. $\alpha_l=\alpha$ which results in $\alpha=N_\text{mda}$.
        Alternatively, the inflation parameters $\alpha_l$ can be self-adaptive based on user-defined criteria~\cite{le2016adaptive}.
        The EnKF-MDA is currently implemented in DAFI using a constant inflation parameter. 
        For linear models, results with different inflation parameters converge to the same solution and only differ in the number of iterations. 
        For nonlinear cases, a large inflation parameter is suggested to damp the correction and alleviate the effects of the model nonlinearity.
        Like for the iterative EnKF, this iterative procedure is done in the inner DAFI loop.

    \subsection{EnRML}
    \label{sec:methods:enrml}
        The ensemble randomized maximum likelihood (EnRML)~\cite{gu2007iterative,chen2012ensemble} method was initially proposed by Gu et al.~\cite{gu2007iterative} for strongly nonlinear systems.
        Randomized maximum likelihood (RML) is a technique that randomizes the likelihood function and converts the maximum a posteriori estimate to a minimization of an objective function which is solved with an optimization technique, e.g., Gauss-Newton algorithm. 
        EnRML introduces the ensemble technique into RML to estimate the complete posterior distribution. 
        Similar to the EnKF-MDA, the EnRML is an iterative method that can damp the change of the nonlinear model with the Gauss-Newton algorithm and thus alleviate the effects of the nonlinearity. 
        It introduces an iteration which is done in the inner loop of DAFI.
        The update scheme of the EnRML method can be formulated as
        \begin{subequations}
            \begin{gather}
                \begin{split}
                \vect{x}_{l+1}^{(j)} = \gamma \vect{x}_{0}^{(j)} +\left(1- \gamma\right) \vect{x}_l^{(j)} - &\gamma C_{\vect{x},0} \left(\mapl{Z}_l^{\prime}\right)^\top \left(\mapl{R}+\left(\mapl{Z}_l^{\prime}\right)^\top C_{\vect{x},0} \mapl{Z}_l^{\prime}\right)^{-1} \\
                &\left(\vect{z}_l^{(j)}-\vect{y}^{(j)}-\mapl{Z}_l^{\prime}\left(\vect{x}_l^{(j)}-\vect{x}_{0}^{(j)}\right)\right) \text{,}
                \end{split} \\
                \vect{z}_l^{(j)} = \mapl{H}\vect{x}_l^{(j)}
            \end{gather} 
            \label{eq:update_EnRML}
        \end{subequations}
        where $\gamma$ controls the step length of the Gauss--Newton update, $\vect{x}_{0}$ and $C_{\vect{x},0}$ are the initial state covariance in one data assimilation window, and $\mapl{Z}_l^{\prime}$ is the sensitivity matrix. 
        The sensitivity matrix is defined by the relationship
        \begin{equation}
            \left[\vect{z}_l-\overline{\vect{z}}_l\right] = \mapl{Z}_l^\prime \left[ \vect{x}_l - \overline{\vect{x}}_l \right] \text{,}
        \end{equation}
        where $\left[\vect{z}_l-\overline{\vect{z}}_l\right]$ and $\left[ \vect{x}_l - \overline{\vect{x}}_l \right]$ denote the matrices of mean subtracted samples. 
        The matrix $\left[\vect{x} - \overline{\vect{x}}\right]$ is non-full, and its inverse is estimated based on singular value decomposition. 
        The step length parameter $0<\gamma\leq1$ can be determined by a standard line search. 
        In this code, it is a user-specified constant. 
        If $\gamma=1$ the iteration performs a full Gauss-Newton update, while if $\gamma=0$ no update is performed. 
        For values between $0$---$1$ the Gauss-Newton update is damped. 
        For strongly nonlinear systems, a small $\gamma$ is recommended to prevent the overcorrections at early iterations. 
        The update scheme assimilates observation data to optimize the state $\mathsf{x}$ iteratively, starting from the initial prior distribution $\mathsf{x}_0$. 
        The iterations are stopped once the convergence criteria or maximum iteration number is reached.

\section{Implementation}
\label{sec:implementation}
\begin{figure}[!htb]
      \centering
      \includegraphics{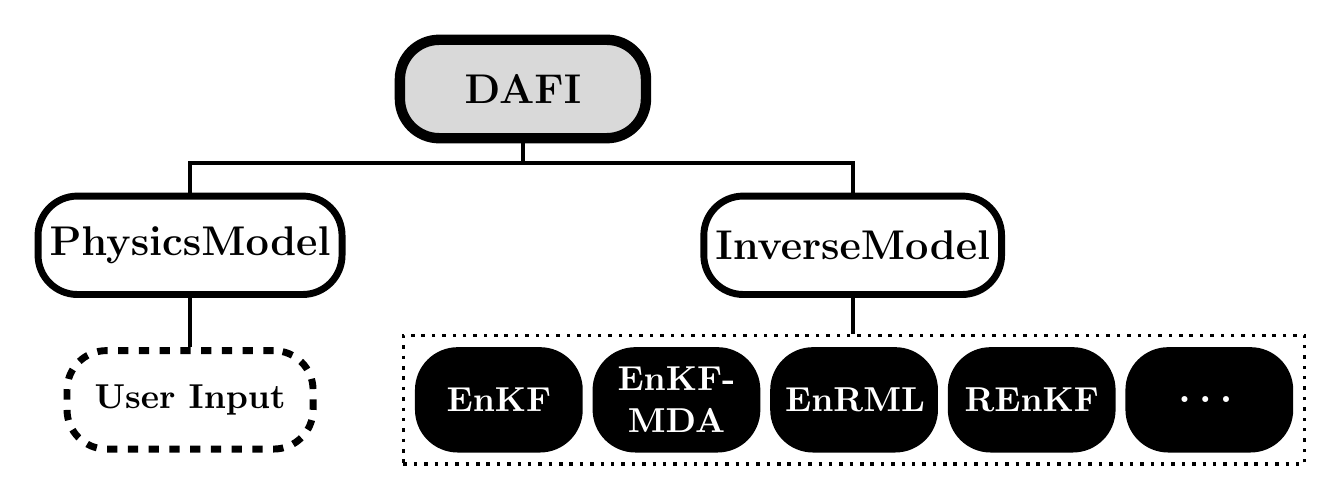}
      \caption{Overview of code structure. DAFI contains two main classes, the PhysicsModel and InverseMethod classes. The physics model is problem-dependent and provided by the user. DAFI contains a library of different inverse methods from which the user selects one. The methods implemented as of the first release are shown.}
      \label{fig:overview}
    \end{figure}

    DAFI is implemented in Python, is available for download from PyPI~\cite{dafi_pypi} (through the \texttt{pip} command) and has online documentation through Read the Docs~\cite{dafi_rtd}.
    The active development repository is hosted in GitHub~\cite{dafi_github}.
    DAFI consists of a Python module and an executable.
    The DAFI module can be loaded with the Python command \texttt{import dafi} and ran with the \texttt{dafi.run(<inputs>)} method.
    The inputs include the name of the inverse method implemented in \texttt{dafi.inverse}, the path to the physics model file, the number of samples, the data assimilation times (outer loop), the maximum number of iterations at each data assimilation time (inner loop), and two dictionaries containing the required inputs for the chosen inverse method and physics model.
    Alternatively the executable can be used to run DAFI from the command line using an input file as \texttt{dafi <inputfile>}.
    DAFI is implemented in an \emph{object-oriented} manner with two main classes: one corresponding to the physics model and another to the inversion method (e.g. EnKF).
    Running DAFI requires an instance of each.
    The inverse method object can be selected as one from the provided library of methods, while the user is required to provide a case specific physics model.
    As such DAFI serves as a robust framework for diverse inverse problems involving fields, with the solution method or problem physics easily exchanged.
    The only user requirement is a physics model that follows a prescribed structure.
    This serves as an API connecting the DAFI solver and the user-specific problem.
    The physics model can be entirely implemented in Python or it can be simply a wrapper for existing solvers such as commercial tools. 
    A simple physics model, implemented completely in Python, is shown in Appendix~\ref{app:code} as an example of writing such a code for a user-specific problem. 
    Figure~\ref{fig:overview} shows an overview of DAFI's structure.
    The following two subsection describe the two classes in more detail.

    \subsection{PhysicsModel Class}
    \label{sec:implementation:physics}
      The physics model is problem-specific and is provided by the user as a class called \texttt{Model} which should be a subclass of the provided \texttt{dafi.PhysicsModel}.
      The required inputs, attributes, and methods for the \texttt{Model} class are briefly described here.
      Several example models and tutorials are provided within the GitHub repository and in the documentation for reference.
      The model is responsible for creating the initial distribution of state samples, propagating a state in time, mapping a state to observation space, and providing the observations and observation errors at each data assimilation time step.
      As such it is an API through which the user provides all the problem specific information.
      The required methods are the following:
      \begin{description}
          \item [\texttt{generate\_ensemble} -] Creates the initial ensemble of state samples. This is a Monte Carlo representation of the prior distribution as discussed in Section~\ref{sec:methods}.
          \item [\texttt{forecast\_to\_time} -] Forecasts the states to the next data assimilation time. This correspond to the dynamic model in Equation~\eqref{eq:dynamic}.
          \item [\texttt{state\_to\_observation} -] Maps the state to observation space. This implements the nonlinear observation operator in Equation~\eqref{eq:Hx}.
          \item [\texttt{get\_obs} -] Provide the observations and observation error at the current data assimilation time step. These are discussed in Section~\ref{sec:general}. 
      \end{description}
      These methods corresponds to different tasks described in Section~\ref{sec:general}.
      Particularly, the dynamic model \texttt{forecast\_to\_time} and the nonlinear observation operator \texttt{state\_to\_observation} are shown in blue in Figure~\ref{fig:filter2} and Algorithm~\ref{alg:DA}. 
      The example code in Appendix~\ref{app:code} shows the physics model for the scalar problem presented in Section~\ref{sec:cases:uq}, showing a simple implementation of each of the methods described above.

    \subsection{InverseMethod Class}
    \label{sec:implementation:inverse}
      The \texttt{InverseMethod} class represents a specific ensemble-based inverse method (e.g. EnKF).
      The different methods discussed in Section~\ref{sec:methods} are implemented as child classes of the \texttt{InverseMethod} class and are available through the \texttt{dafi.inverse} module.
      For most applications the user simply chooses one of the implemented methods.
      The \texttt{InverseMethod} class implements all the details of the general problem described in Section~\ref{sec:general}.
      This includes the outer time loop, inner iteration loop, and stopping criteria.
      Missing from the \texttt{InverseMethod} class however is any specific analysis scheme (i.e. filter $\mapl{A}$), which is implemented through the \texttt{analysis()} method of child classes.
      The specific methods (e.g. EnKF, EnRML) are implemented as child classes of \texttt{InverseMethod} and inherit the \texttt{solve} method but implement their own \texttt{analysis} method.
      Creating such a child class is also the recommended approach for creating new inverse methods.
      The object-oriented and open source nature of DAFI makes it straightforward to implement new inverse methods.

\section{Fields}
\label{sec:fields}
    DAFI was built with fields in mind.
    Its core implementation however has no concept of fields, just state and observation vectors.
    When a problem has a state vector that includes discretized fields it is up to the user to account for this when creating the problem-specific physics model.
    DAFI includes two python modules to aid in creating the physics model: \texttt{dafi.random\_fields} and \texttt{dafi.random\_fields.covariance}.
    The \texttt{dafi.random\_fields} module includes functions for tasks such as calculating the Karhunen-Loève decomposition, calculating the norm of a field, and calculating the projections of a field onto a set of basis.
    It also includes classes representing stochastic processes with classes with methods to do tasks such as generating samples. 
    Currently two such classes for Gaussian and lognormal processes are implemented. 
    The \texttt{dafi.random\_fields.covariance} submodule contains functions to perform tasks related to covariance and correlation matrices, including creating covariance matrices based on some implemented covariance kernels such as the square exponential kernel.
    The relevant theory is presented in this section and the physics models for the tutorials serve as simple examples on how to use these modules when creating a user-specific physics model.

    When dealing with fields, solving the dynamic and forward models requires discretizing the domain.
    All fields are then represented by the finite vector of their values at each cell.
    These discretized values would then be included in the state vector.
    From the point of view of the DAFI code, there is no distinction between state vectors that include discretized fields and ones that do not.
    The notion of a physically continuous field, however, does come into play in several user specified inputs.
    Specifically, the appropriate norm for a field quantity at time $t_i$ is defined by the inner product between two fields $f_1$ and $f_2$ as
    \begin{subequations}
        \begin{gather}
            \left<f_1(\bm{\xi}),f_2(\bm{\xi})\right> = \int_\Omega f_1(\bm{\xi}) f_2(\bm{\xi})\mathrm{d}\Omega \approx \sum_{j=1}^{N_\Omega} f_1(\bm{\xi}_j) f_2(\bm{\xi}_j) \Delta\Omega_j \text{,} \\
            \lVert f(\bm{\xi}) \rVert^2 = \left< f(\bm{\xi}),f(\bm{\xi}) \right> \text{,}
        \end{gather}
        \label{eq:norm}
    \end{subequations}
    where $\Omega$ is the spatial domain, $N_\Omega$ is the number of cells in the discretization, $\bm{\xi}_j$ is the coordinate of the $j$\textsuperscript{th} cell's center, and $\Delta\Omega_j$ is the $j$\textsuperscript{th} cell's volume.
    For simplicity of notation the constant $t_i$ argument for all fields was omitted.
    Note that when the field is discretized, calculating the inner product and norm requires taking into account the volume of each cell.
    In practice this is enforced by an appropriate choice of the weight matrix in the L2 norm.
    The norm of a vector $\vect{x}$ with given weight matrix $W$ is given by
    \begin{equation}
        \lVert \vect{x} \rVert^2_W = \vect{x}^\top W \vect{x} \text{.}
        \label{eq:discretenorm}
    \end{equation}
    For the case where the vector consists of only a discretized field the weight matrix is a diagonal matrix with $W_{jj}=\Delta\Omega_j$.

    Another place where the concept of a continuous field comes into play is in the specification of a covariance matrix for the prior state distribution.
    This is required in the \texttt{PhysicsModel.generate\_ensemble} method.
    If the state vector includes discretized fields, the covariance matrix should reflect the physical correlation of these fields.
    For example, a common covariance kernel for spatial correlation is the squared exponential kernel, where the covariance between the values of the field at two spatial locations is given as a function of the distance between the two locations.
    Using the squared exponential kernel the covariance matrix $C$ for a vector $\vect{x}$ consisting of a single discretized field is given by
    \begin{equation}
        C_{j,k} = \mathcal{C}(\bm{\xi_j},\bm{\xi_k}) = \sigma^2 \exp\left( -\frac{1}{2} \frac{\lVert \bm{\xi}_j-\bm{\xi}_k \rVert^2}{l^2} \right) \text{,}
        \label{eq:covariance}
    \end{equation}
    where the parameters $\sigma^2$ and $l$ are the variance and length scale, respectively, and $\mathcal{C}$ is the continuous covariance kernel.
    More complex examples might use different kernels for the spatial covariance, such as a periodic kernel.
    Wu et al.~\cite{wu2019physics} proposed a method of incorporating known physical constraints for input fields through choice of covariance matrix.
    Similarly Michelén~Ströfer et al.~\cite{michelenstrofer2019enforcing} used the choice of covariance matrix to enforce boundary conditions on the input fields.

    Using the Karhunen-Loève (KL) decomposition, a random field can be represented as an infinite linear combination of orthogonal basis functions, referred to as modes, where the coefficients of the linear combinations are random variables.
    In the case of a Gaussian process the coefficients are identically distributed independent random variables with standard normal distribution.
    That is, the random field at a given time $t_i$ can be written as
    \begin{equation}
        f(\bm{\xi}) - \overline{f}(\bm{\xi}) = \sum_{j=1}^\infty \omega_{j} \phi_{j}(\bm{\xi}) \approx \sum_{j=1}^{N_\Omega} \omega_{j} \phi_{j}(\bm{\xi}) \approx \sum_{j=1}^{N_m} \omega_{j} \phi_{j}(\bm{\xi}) \text{,}
        \label{eq:kl}
    \end{equation}
    where $\overline{f}$ is the mean field, $\omega_{j}$ are the random coefficients and $\phi_{j}(\bm{\xi})$ are the modes.
    Here again the constant $t_i$ argument for all fields was omitted and it is understood that the modes and coefficients correspond to the field at that time.
    The first approximation corresponds to the discretization of the domain and the second corresponds to an approximation using only $N_m<N_\Omega$ modes.
    The first few modes contain most of the variance and it is typical to represent the field with a subset of $N_m$ modes chosen to cover some percentage (e.g. $99\%$) of the variance.
    For a Gaussian process each coefficient has a standard normal distribution $\omega_j \sim \mathcal{N}(0,1)$.
    The user can use the KL decomposition to create the initial set of samples as an alternative to the Cholesky decomposition.
    Additionally, the KL decomposition can be used as a reduced order model where the coefficients $\{\omega_j\}_{j=1}^{N_m}$ are the state rather than the discretized values of the field~\cite{xiao_quantifying_2016}. 

    The KL modes are given by the eigendecomposition of the covariance kernel.
    For a continuous covariance kernel $\mathcal{C}(\bm{\xi_1},\bm{\xi_2})$ the eigenvalues $\lambda_k$ and eigenfunctions $e_k(\bm{\xi_1})$ are obtained by solving the associated Fredholm integral equation
    \begin{equation}
        \int_\Omega \mathcal{C}(\bm{\xi_1},\bm{\xi_2})e_k(\bm{\xi_1})\mathrm{d}\bm{\xi_1} = \lambda_k e_k(\bm{\xi_2}) \text{,}
        \label{eq:klintapp}
    \end{equation}
    and sorting the eigenpairs based on largest eigenvalues with $\lambda_1$ corresponding to the largest eigenvalue.
    With unit eigenfunctions $\hat{e}_k(\bm{\xi_1})$, the modes are given by
    \begin{equation}
        \phi_k(\bm{\xi}) = \sqrt{\lambda_k}\hat{e}_k(\bm{\xi}) \text{,}
        \label{eq:klmodes}
    \end{equation}
    where the normalized eigenfunctions describe the mode shapes and the eigenvalues are the variance associated with each mode. 
    For the discretized case the Fredholm integral equation in Equation~\eqref{eq:klintapp} becomes (See Appendix~\ref{app:kl})
    \begin{equation}
        \left( C W \right)\left(e_k\right) = \lambda_k e_k \text{,}
    \end{equation}
    where $C$ and $e_k$ are now the covariance matrix and an eigenvector, respectively.
    The weight matrix $W$ is diagonal with entries $W_{ii}=\Delta\!\Omega_i$, where $\Delta\!\Omega_i$ is the volume of the $i$\textsuperscript{th} cell in the discretization.
    The modes are then given by 
    \begin{equation}
        \phi_k = \sqrt{\lambda_k}\hat{e}_k \text{.}
    \end{equation}
    Note that normalizing the eigenfunctions and eigenvectors is done with the L2 norm for a field given in Equation~\eqref{eq:norm} and Equation~\eqref{eq:discretenorm}. 
    An alternative representation of the discrete problem is also presented in Appendix~\ref{app:kl}.

    \subsection{Coupling to Third Party Software: OpenFOAM Example}
    \label{sec:fields:openfoam}
      The physics solver for most applications is likely to be either an in-house or commercial third party software. 
      For instance DAFI has mostly been used to solve the RANS equations~\cite{xiao_quantifying_2016,zhang2019regularization,michelenstrofer2019enforcing} using the open source finite volume tool OpenFOAM~\cite{openfoam}. 
      DAFI includes a tutorial that couples OpenFOAM and DAFI to solve the RANS equations.
      This serves as a general example on how to couple DAFI with other third party physics solvers, and as a specific example for developing other OpenFOAM-based physics models.
      The tutorial solves the RANS field inversion problem where the eddy viscosity is inferred from observations of the velocity for a two-dimensional flow over infinite periodic hills (see Figure~\ref{fig:phills} later). 
      The physics model \texttt{nutfoam.py} uses OpenFOAM for solving the RANS equations (forward problem). 
      It serves as a more complex example of writing a physics model, using the \texttt{fields} and \texttt{covariance} modules, and coupling to third party solvers. 
      To facilitate developing other OpenFOAM-based solvers, DAFI also includes the module \texttt{dafi.foam} for OpenFOAM file input/output (I/O) operations.
      
      In addition to OpenFOAM DAFI is being coupled to the open-source NHWave~\cite{ma2012shock} software by the developers at Old Dominion University. 
      As an open source project, third party coupling with other software by the community are welcomed and could be included in future releases of the code. 
    
    \subsection{Example: Working with Fields}
    \label{sec:fields:example}
\begin{figure}[!htb]
        \begin{subfigure}{0.49\linewidth}
            \centering
            \includegraphics[width=\linewidth]{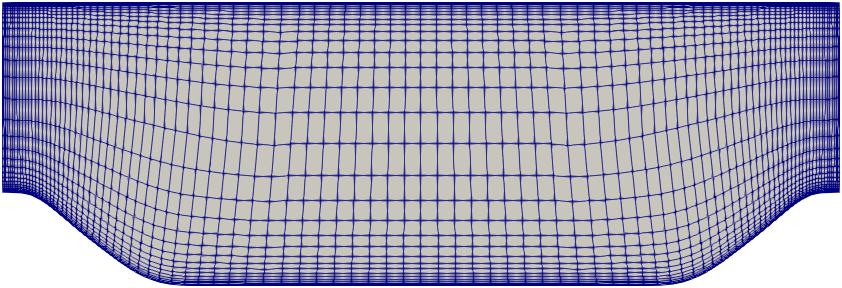}
            \caption{Mesh}
            \label{fig:phills:mesh}
        \end{subfigure}
        \begin{subfigure}{0.49\linewidth}
            \centering
            \includegraphics[width=\linewidth]{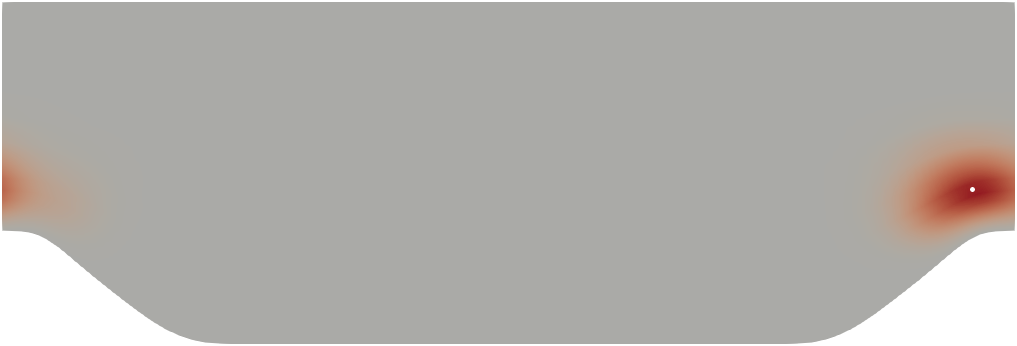}
            \caption{Covariance of a field value at point $\bm{\xi}_i$ (white)}
            \label{fig:phills:cov}
        \end{subfigure} 
        \caption{Periodic hills case as example of working with fields. Panel (a) shows the computational mesh consisting of $3000$ cells. The top and bottom boundaries are walls, while the left and right boundaries are periodic. Panel (b) shows the covariance between a field value at a single point $\bm{\xi}_i$, shown in white, and the field value at all other points in the domain.}
        \label{fig:phills}
    \end{figure}

    DAFI handles fields with arbitrary boundary shapes and discretization, i.e. it is not limited to constant-spacing structured meshes. 
    This section shows an example of performing some field operations using DAFI. 
    The case corresponds to the infinite periodic hills~\cite{breuer2009flow} example available in the tutorials, where a single hill is modeled using periodic boundary conditions. 
    The top boundary is a wall.  
    The mesh used is shown in Figure~\ref{fig:phills:mesh} and consists of $3000$ cells. 
    An important task is creating a covariance matrix for the input field (state to be inferred). 
    The square exponential kernel is chosen for the vertical direction, and a periodic kernel is used for the periodic direction. 
    The covariance kernel is then modified \cite{michelenstrofer2019enforcing} to enforce zero covariance at the wall boundaries which have fixed-value Dirichlet boundary condition.  
    The covariance between the field value at a single cell the field value at all other cells in the domain is shown in Figure~\ref{fig:phills:cov}. 
    This corresponds to a single row (or column) of the covariance matrix. 
    It can be seen that the point is highly correlated to its immediate neighbors (even across the periodic boundary) and then quickly becomes uncorrelated with the rest of the domain. 
    The covariance also goes to zero towards the wall. 
    
\begin{figure}[!htb]
        \centering
        \begin{subfigure}{0.49\linewidth}
            \centering
            \includegraphics[width=0.65\linewidth]{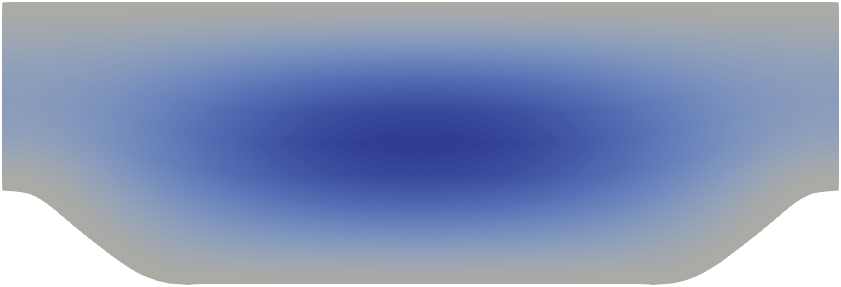}\\
            \includegraphics[width=0.65\linewidth]{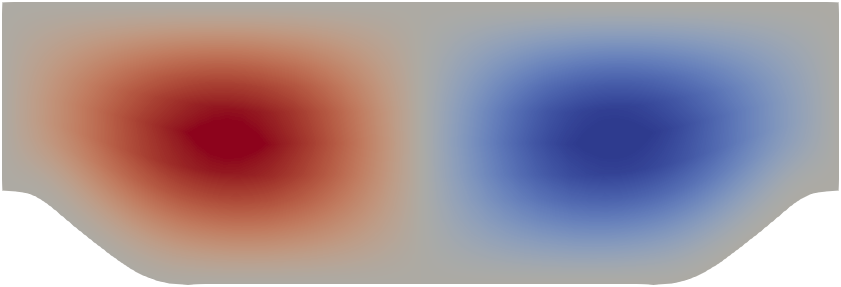}\\
            \includegraphics[width=0.65\linewidth]{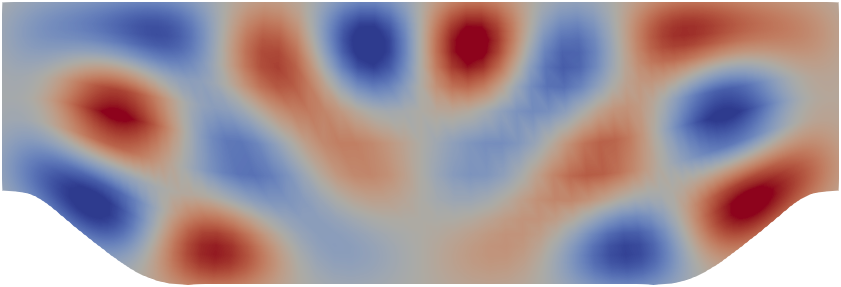}
            \caption{Select KL mode shapes}
            \label{fig:phills_modes:modes}
        \end{subfigure}
        \begin{subfigure}{0.49\linewidth}
            \centering
            \includegraphics[width=\linewidth]{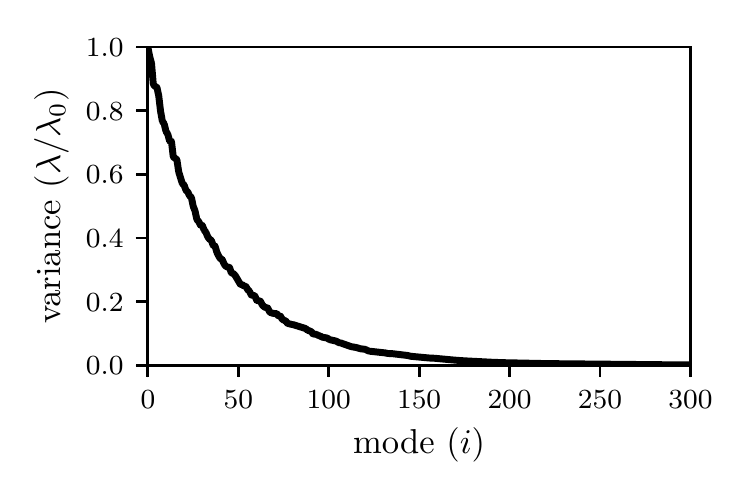}
            \caption{Variance}
            \label{fig:phills_modes:var}
        \end{subfigure} 
        \caption{KL decomposition of covariance kernel for the periodic hills example of working with fields. Panel (a) shows mode shapes (normalized eigenvectors) for modes 1, 2, and 25. It can be seen that higher modes account for variations at smaller length scales. Panel (b) shows the variance associated with each mode. It can be seen that the first $200$ modes capture almost all the variance.}
        \label{fig:phills_modes}
    \end{figure}

    A follow on, optional, task is calculating the discrete modes for the KL decomposition of the covariance matrix.  
    This decomposition can be used to generate samples from the random field or truncated and used as a reduced order model, as shown in Equation~\eqref{eq:kl}. 
    The mode shapes (normalized eigenvectors) for a few modes, and the variance (eigenvalues) of each mode are shown in Figure~\ref{fig:phills_modes:modes} and~ Figure~\ref{fig:phills_modes:var} respectively. 
    The initial modes are seen to capture variance at large length scales, while the latter modes account for smaller length scales.
    To generate the prior samples, the modes would be weighted by the coefficients $\omega_i$ and added to the mean field following Equation~\eqref{eq:kl}. 
    For a Gaussian process, which the prior distribution is typically assumed to be, the coefficients are sampled from independent standard normal distributions (i.i.d.), i.e. $\omega_i\sim\mathcal{N}(0,1)$.
    For reduced order modeling, it can be seen in Figure~\ref{fig:phills_modes:var} that the first $200$ modes capture virtually all the variance. 
    Using this reduced set of modes to represent the state reduces the dimensionality of the state space from $3000$, corresponding to the number of cells, to $200$, corresponding to the number of retained KL modes. 
    The state would now consist of the $200$ coefficients $\omega_i$. 

\section{Test Cases}
\label{sec:cases}
  Several test cases are presented in this section to showcase the range of problems DAFI is meant to solve.
  The first test case solves the classical filtering problem for a chaotic dynamic system governed by the Lorenz equations using the EnKF. 
  In this case the state consists of three scalar quantities related through three governing equations. This example, while simple, serves as an introduction to data assimilation and the filtering procedure. 
  The second test case is a field inversion problem where the diffusivity (input) field is inferred from observations of the output field of the diffusion equation. 
  This is first solved with the iterative EnKF which leads to improvement in the output field but to an input field with undesirable characteristics. 
  This is possible because of the ill-posed nature of field inversion problems, where typically many input fields can result in similar output fields. 
  The third test case uses the same diffusion field inversion problem but uses the regularized EnKF to enforce smoothness in the input field. 
  This leads to similar improvements in the output field but with better behaved input field. 
  The fourth test case shows the results, using DAFI and reproduced with permission from Michelén~Ströfer et al.~\cite{michelenstrofer2019enforcing}, of field inversion problem using the RANS equations where the eddy viscosity field is inferred from velocity observations. This constitutes an example of a problem of practical importance and of using a more complex, non-trivial geometry.
  Finally the fifth test case solves a simple scalar inverse problem described by algebraic equations, but focuses on the ability of doing uncertainty quantification with DAFI. 
  Different methods are used and the advantage of either the EnKF-MDA or the EnRML over the EnKF for uncertainty quantification is demonstrated. 
  These test cases collectively give an overview of DAFI's capabilities and are available in the GitHub repository.

    \subsection{Data Assimilation: Lorenz Equations}
    \label{sec:cases:lorenz}
        The Lorenz equations, first introduced by Edward Lorenz in 1963~\cite{lorenz}, are a well known system of chaotic ordinary differential equations. 
        The system can be used to model atmospheric convection and contains three states governed by the following three coupled PDE
        \begin{align}
            \frac{d\vecti{x}_1}{dt} = \sigma(\vecti{x}_2-\vecti{x}_1) \\
            \frac{d\vecti{x}_2}{dt} = \rho \vecti{x}_1 - \vecti{x}_2 -\vecti{x}_1\vecti{x}_3 \\
            \frac{d\vecti{x}_3}{dt} = \vecti{x}_1\vecti{x}_2 - \beta \vecti{x}_3 \text{,}
        \end{align}
        where $\sigma, \rho$, and $\beta$ are real positive model parameters. 
        These equations constitute the dynamic model, which is used to propagate a given state in time. 
        The state $\vecti{x}_1$ is proportional to the rate of convection, and the states $x_2$ and $\vecti{x}_3$ are proportional to the horizontal and vertical temperature differences, respectively~\cite{lorenz}. 
        The Lorenz equations are chaotic and small deviations in initial conditions can quickly result in large differences in the state. 
        The chaotic behavior can be controlled by using data assimilation to correct the states whenever observations are available. 
        
\begin{figure}[!htb]
            \begin{subfigure}{\linewidth}
        		\centering
        		\includegraphics[scale=0.94]{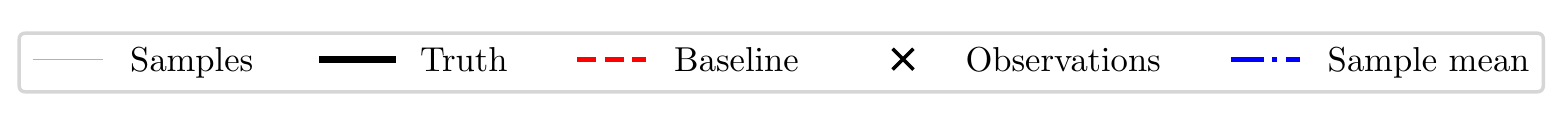}
        	\end{subfigure}
        	
        	\begin{subfigure}{0.49\linewidth}
        		\centering
        		\includegraphics[scale=0.95]{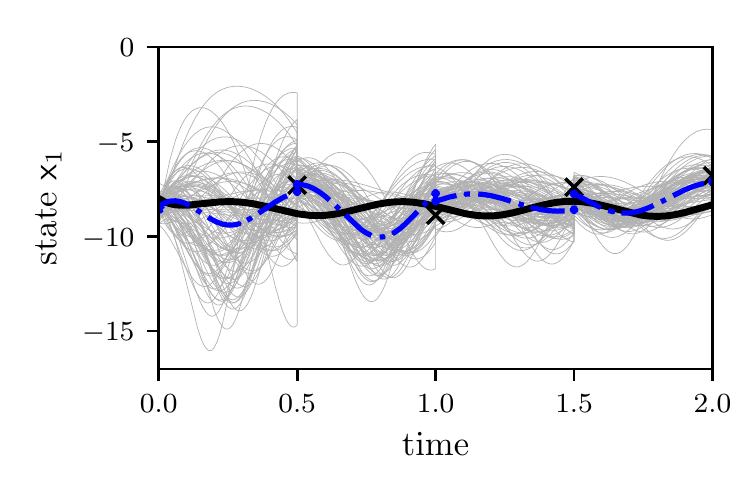}
                \caption{State $\vecti{x}_1$, $0\leq t \leq 2$}
                \label{fig:x102}
        	\end{subfigure}
            \hfill
        	\begin{subfigure}{0.49\linewidth}
        		\centering
        		\includegraphics[scale=0.95]{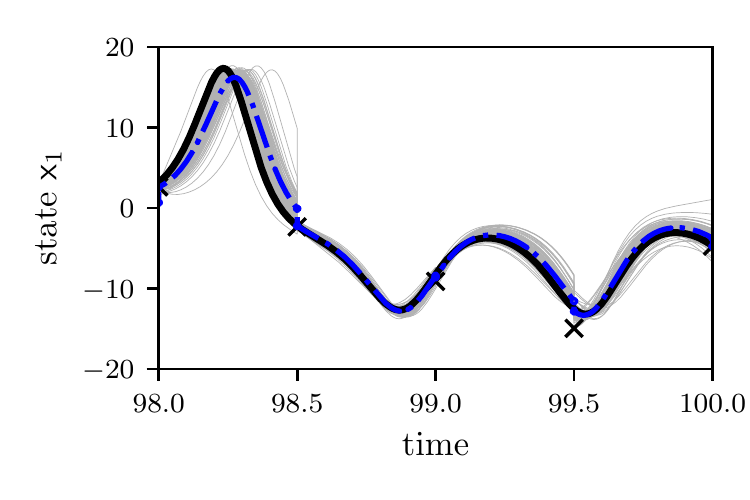}
                \caption{State $\vecti{x}_1$, $98\leq t \leq 100$}
                \label{fig:x198}
        	\end{subfigure}
        
        	\begin{subfigure}{0.49\linewidth}
        		\centering
        		\includegraphics[scale=0.95]{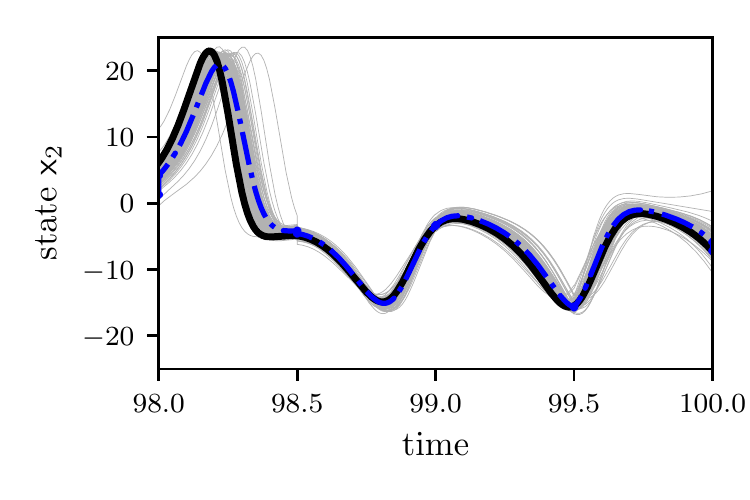}
                \caption{State $\vecti{x}_2$, $98\leq t \leq 100$}
                \label{fig:x298}
        	\end{subfigure}%
        	\begin{subfigure}{0.5\linewidth}
        		\centering
        		\includegraphics[scale=0.95]{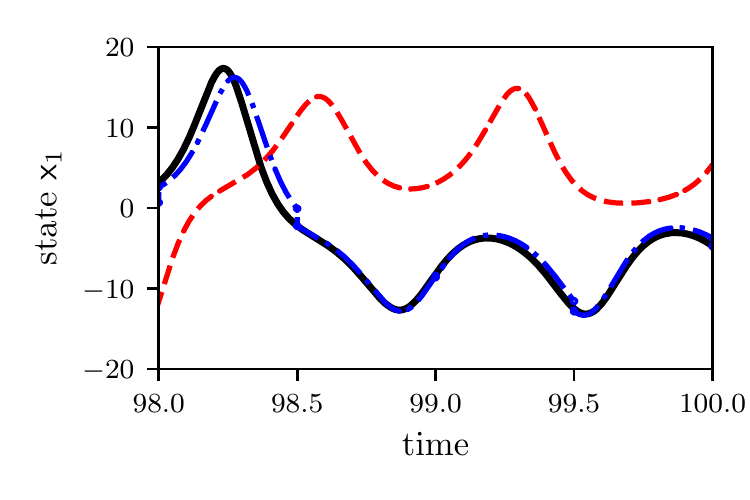}
                \caption{State $\vecti{x}_1$, $98\leq t \leq 100$}
                \label{fig:x1comp}
        	\end{subfigure}%

        	\caption{State estimation results for the Lorenz data assimilation test case. Panel~(a) shows state $\vecti{x}_1$ at the start of the process, where the predictions can be seen to disagree with the truth. Panel~(b) shows the same after the data assimilation has gone on for long time, and the ensemble mean can be seen to be very close to the truth with small uncertainty. Panel~(c) shows the state $\vecti{x}_2$, which shows that improvements are made even for unobserved states. Panel (d) compares the data assimilation solution and the baseline solution (no data assimilation) to the truth.}
        	\label{fig:lorenz}
        \end{figure}

        The baseline initial condition is chosen as $\vecti{x}_{1}(0)=-8.5, \vecti{x}_{2}(0)=-7.0, \vecti{x}_{3}(0)=27.0$ which is slightly off from the true initial conditions $\vecti{x}_{1}(0)=-8.0, \vecti{x}_{2}(0)=-9.0, \vecti{x}_{3}(0)=28.0$. 
        Additionally, the parameters $\beta$ and $\sigma$ are taken as correct, but the parameter $\rho$ is considered uncertain. 
        The state augmentation procedure is used and $\rho$ is added to the state, which is now 
        \begin{equation}
              \vect{x} = \begin{bmatrix} \vecti{x}_1\\ \vecti{x}_2\\ \vecti{x}_3\\ \rho \end{bmatrix} \text{.}
        \end{equation}
        The parameter $\rho$ is chosen as $29$ whereas the true value is taken as $28$.
        As can be seen in Figure~\ref{fig:lorenz} the baseline solution quickly diverges from the truth. 
        We use the true solution to generate synthetic observations, which consists of observations of state $\vecti{x}_1$ and $\vecti{x}_3$ every half second. 
        This correspond to the observation operator 
        \begin{equation}
              \mapl{H} =  \begin{bmatrix}1 & 0 & 0 & 0  \\ 0 & 0 & 1 & 0 \end{bmatrix} \text{.}
        \end{equation}
        These observations each have an uncorrelated observation error (variance) based on relative and absolute standard deviations of $r=0.1$ and $a=0.05$, as
        \begin{equation}
            C_{\vect{y},i,i} = \left(\vecti{y}^*_i  r + a \right)^2 \label{eq:synth_obs} \text{,}
        \end{equation}
        for $i\in{1,2}$, where $\vect{y}^*$ is the synthetic truth.
        Finally, the confidence in our initial values needs to be specified to create the prior distribution. 
        The prior is taken as a multivariate Gaussian distribution with mean equal to the baseline values listed above and uncorrelated covariance matrix
        \begin{equation}
            C_{\vect{x}} = \operatorname{diag}(0.4, 2.0, 1.4, 4.0) \text{.}
        \end{equation}
        
\begin{figure}[!htb]
            \centering
            \includegraphics[width=0.6\textwidth]{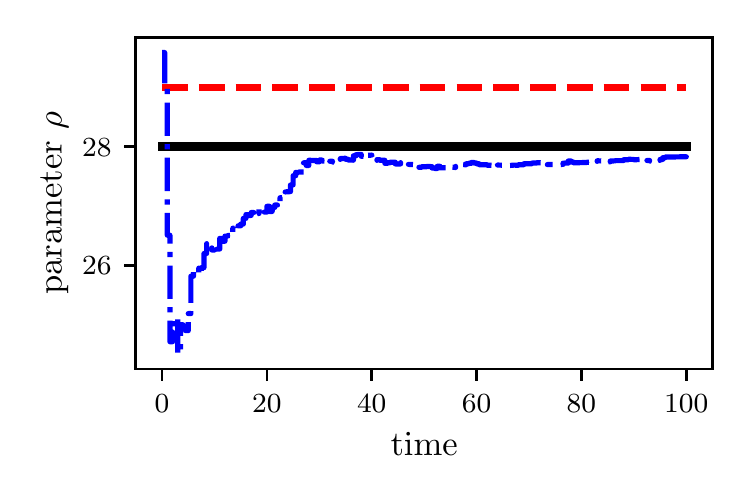}
            \caption{Parameter estimation results for the Lorenz data assimilation test case. Starting from a baseline solution (red/light grey, dashed line), the inferred sample mean (blue/dark grey, dash-dotted line) for the parameter $\rho$ gradually approaches the ground truth (black, solid line) as more data is assimilated. The legend here follows that for Figure~\ref{fig:lorenz}.}
            \label{fig:rho}
        \end{figure}

        With the prior distribution specified, the dynamic model available, and the observations and their error available, the EnKF is used to correct the state each time an observation is made. 
        The results of EnKF with $100$ samples are shown in Figure~\ref{fig:lorenz}. 
        The figure shows two states, $\vecti{x}_1$ and $\vecti{x}_2$, at two different time windows, one at the beginning of the process and the other after the process has been going on for a while. 
        Each sample in the ensemble is propagated with the dynamic model. 
        When an observation becomes available each sample is updated using the observation and ensemble statistics. 
        It can be seen that at the beginning the state distribution has a mean far from the truth with very large covariance. 
        After the data assimilation process has gone on for a while the mean gets very close to the truth and the uncertainty is greatly reduced. 
        This is true for all states, even $\vecti{x}_2$ which is not directly observed. 
        For reference the baseline solution, that is the solution with the initial values with no data assimilation, is also shown. 
        Figure~\ref{fig:rho} shows the inferred value of the parameter $\rho$ which in general gets closer to the truth as more data is assimilated. 
        The inferred parameter $\rho$ initially varies rapidly but quickly stabilizes as more data is assimilated.
        
    \subsection{Field Inversion: Diffusion Equation}
    \label{sec:cases:diffusion}
        The one-dimensional diffusion equation with homogeneous boundary conditions is used in this test case. 
        The diffusion equation is given as 
        \begin{subequations}
            \begin{gather}
            -\frac{d}{d \xi_1}\left(\mu \frac{d u}{d \xi_1}\right)=f(\xi_1) \text{,} \\
            u(0) = u(L) = 0 \text{,}
            \end{gather}
            \label{eqn:diffusion}
        \end{subequations}
        where $u$ is the quantity being diffused (e.g. heat), $\mu$ is the diffusivity, $\xi_1$ is the spatial coordinate along the domain (e.g. a finite rod), $L$ is the domain length, and $f$ is a source term. 
        The problem consists of inferring the non-constant diffusivity field $\mu$ from sparse observations of the output field $u$. 
        The diffusion equations can be seen as a forward model that propagates the state (input field $\mu$) to the observable field (output field) $u$. 
        The forward model is solved using a central finite difference scheme. 
        The source term considered is $f(\xi_1)= sin(0.2\pi \xi_1)$.
        The synthetic truth is created using the first three modes each with coefficient $\omega_i=1$.
        The observations are created from the synthetic truth output field $u$ at $\xi_1/L\in\{0.25, 0.5, 0.75\}$ each with uncorrelated variance based on relative and absolute standard deviations of $0.1$ and $0.0001$. 
        The observation operator then consists of the forward model followed by selection of the three points. 
        
        The domain is discretized into $100$ equally spaced cells and the state could consist of the discretized diffusivity field, however it will be modified to enforce positivity and to showcase dimensionality reduction using KL modes.
        The diffusivity field is physically constraint to be positive, and this is enforced by inferring the logarithm of diffusivity $\text{log}[\mu/\mu_0]$ rather than the diffusivity directly. 
        The baseline solution $\mu_0$ corresponds to our initial guess or prior belief (that is, before seeing the observation data) and was chosen as a constant field. 
        The prior distribution is taken as a Gaussian process with zero mean and specified covariance $C_{\vect{x}}$, that is 
        \begin{equation}
            \text{log}[\mu/\mu_0] \sim \mathcal{GP}(0,C_{\vect{x}}) \text{.}
        \end{equation} 
        This formulation makes diffusivity have a lognormal distribution with median $\mu_0$, and the state being inferred corresponds to the exponent in the multiplicative term $\mu=\mu_0\exp{(x)}$. 
        For the covariance a square exponential kernel is used with a standard deviation of $\sigma_{\text{p}}=5.0$ and length scale $l/L=0.02$. 
        While $100$ states is manageable, for more complex cases involving fields the size of the discretization can become unmanageable.  
        The KL modes can be used for dimensionality reduction as discussed in Section~\ref{sec:fields}. 
        With a choice of the $m$ modes with largest variance, the field can be represented as 
        \begin{equation}
            \log \left[\mu / \mu_{0}\right]=\sum_{i=1}^{m} \omega_{i} \sqrt{\lambda_{i}} \phi_{i} \text{,}
        \end{equation}
        where the variance $\lambda_i$ and mode shapes $\phi_i$ are obtained from the eigendecomposition of the covariance matrix as discussed in Section~\ref{sec:fields}. 
        The state vector now consists of the $m$ coefficients $\omega_i$ and we use $m=15$.
        
        \begin{figure}[!htb]
            \begin{subfigure}{1.0\linewidth}
        		\centering
        		\includegraphics[scale=0.94]{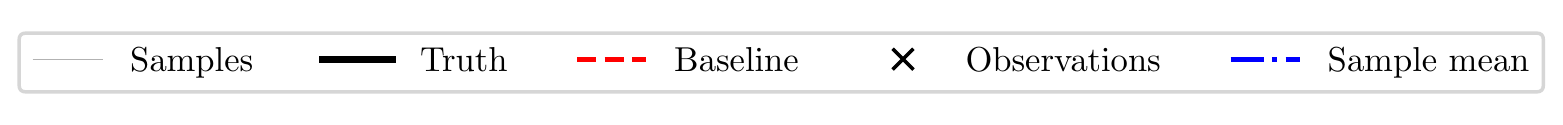}
        	\end{subfigure}
        	
        	\begin{subfigure}{0.49\linewidth}
        		\centering
        		\includegraphics[scale=0.95]{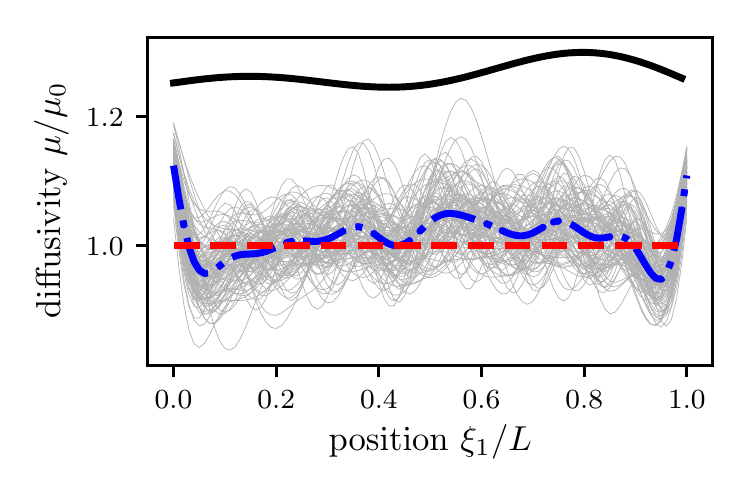}
                \caption{Prior of input diffusivity field}
                \label{fig:enkf_mu_prior}
        	\end{subfigure}
        	\begin{subfigure}{0.49\linewidth}
        		\centering
        		\includegraphics[scale=0.95]{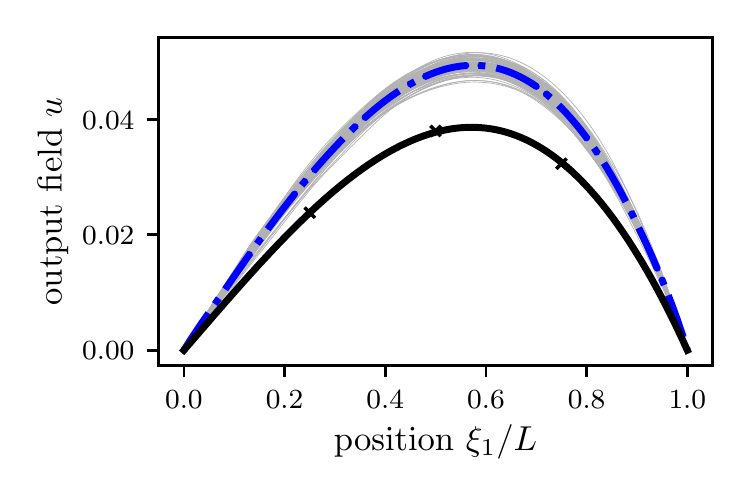}
                \caption{Prior of output field}
                \label{fig:enkf_u_prior}
        	\end{subfigure}
        
            \begin{subfigure}{0.49\linewidth}
        		\centering
        		\includegraphics[scale=0.95]{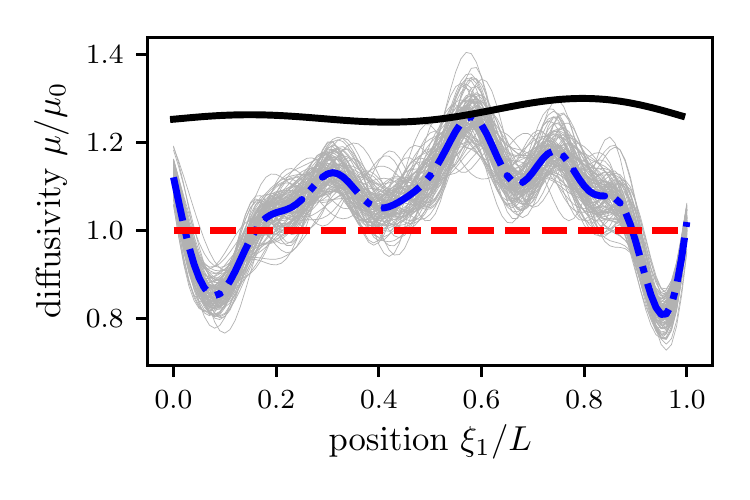}
                \caption{Posterior of input diffusivity field}
                \label{fig:enkf_mu_post}
        	\end{subfigure}
        	\begin{subfigure}{0.49\linewidth}
        		\centering
        		\includegraphics[scale=0.95]{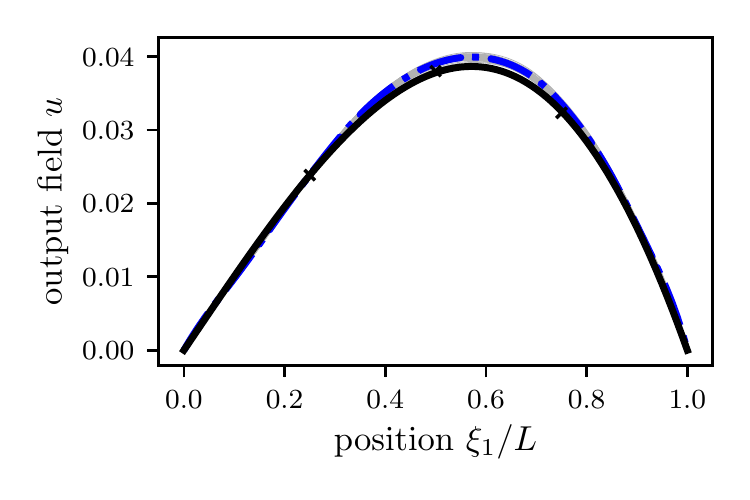}
                \caption{Posterior of output field}
                \label{fig:enkf_u_post}
        	\end{subfigure}
        	\caption{Results for the diffusion equation field inversion test case without regularization. With sparse observations of the output field the inferred output field is greatly improved. The input field however is significantly more challenging to infer correctly since the field inversion problem is ill-posed.}
        	\label{fig:diffusion}
        \end{figure}
        Results of the iterative EnKF using $100$ iterations and $100$ samples are shown in Figure~\ref{fig:diffusion}.  
        The inferred output field $u$ is greatly improved by incorporating the data, but the input field does not approach the true solution. 
        This is because the problem is ill-posed, with many possible input fields resulting in good agreement with the observations. 
        The next subsection solves the same problem using regularization to enforce known or desired properties of the input field.

    \subsection{Regularized Field Inversion}
    \label{sec:cases:regularized}
        The field inversion problem in Section~\ref{sec:cases:diffusion} did not result in an input field close to the truth. 
        This is due to the ill-possedness of the field inversion problem. 
        We will now show how to use the regularized EnKF discussed in Section~\ref{sec:methods:renkf} to enforce desired properties of the input field and alleviate the ill-possedness. 
        Specifically, we notice that the inferred diffusivity field is very jagged and we have knowledge (e.g. expert domain-specific knowledge) that it should be more smooth. 
        As such we will prefer to only use the first three modes if possible and will penalize the use of any additional modes, with the penalty function 
        \begin{equation}
            \mathcal{G}[\omega] = \omega_i \text{,}
        \end{equation}
        and weighting matrix
        \begin{equation}
            W = \lambda\ \operatorname{diag}(0,0,0,1,...,1)\text{,}
        \end{equation}
        where $\lambda$ is regularization strength parameter. 
        This does not prevent the inference from utilizing higher modes, if using them does improve the agreement with observations, but among equally good solutions it prefers ones that use at most the first three modes. 
        
         \begin{figure}[!htb]
    		\centering
    		\includegraphics{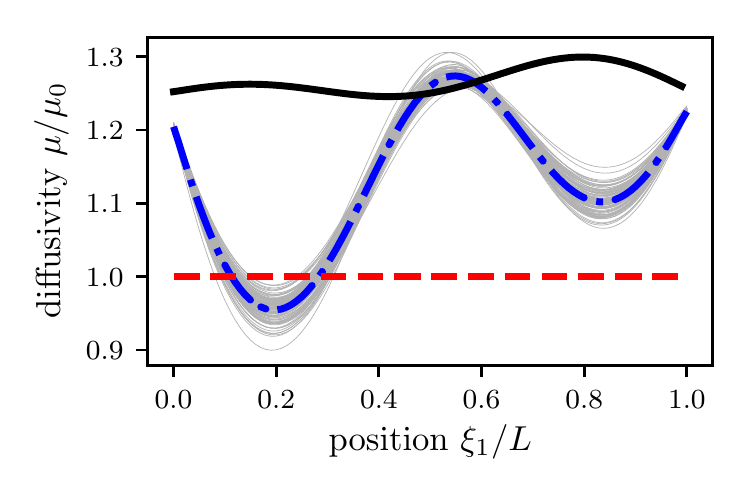}
    		\includegraphics{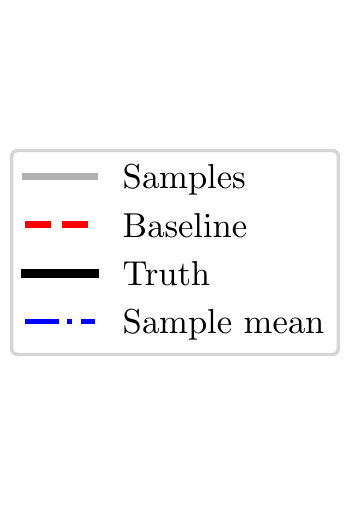}
    	\caption{Results for the regularized field inversion test case for the diffusion equation. While the inferred field is still far from the truth it displays the requested behavior, namely smoothness at a longer length scale.}
    	\label{fig:renkf}
        \end{figure}
        
     The results of the field inversion using the regularized EnKF and a regularization coefficient of $\lambda=10^6$ are shown in Figure~\ref{fig:renkf}. 
     The regularization coefficient used a ramp-up $\tanh$ function over the first $10$ steps to decrease the effect of regularization during the first few iterations. 
     The prior distributions are identical as before and the posterior of the output field displays the same improvement as with the EnKF (Figure~\ref{fig:enkf_u_post}), and therefore those results are omitted.
     The inferred input diffusivity field is still not close to the truth, but it satisfies the desired smoothness. 
     This means that the problem is still ill-posed and further regularization is required. 
     The results show the effectiveness of the method in enforcing prior knowledge. 
     Effectively the space of possible solutions was reduced to those satisfying the desired smoothness property.

    \subsection{Field Inversion: RANS Equations}
    \label{sec:cases:hills}
\begin{figure}[!htb]
        \centering
        \includegraphics[trim=0 0.1in 0 0.1in,clip,scale=0.98]{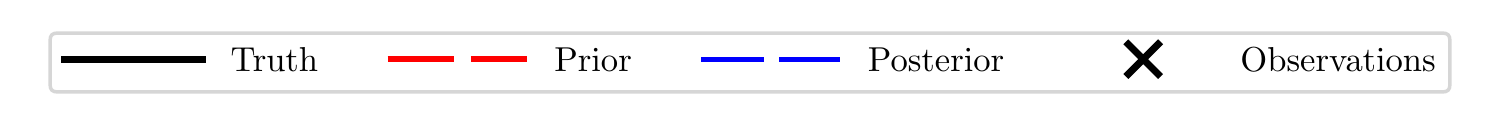}
        
        \begin{subfigure}[b]{0.49\linewidth}
          \centering
            \includegraphics[trim=0.1in 0.15in 0.1in 0.3in,clip,scale=0.96]{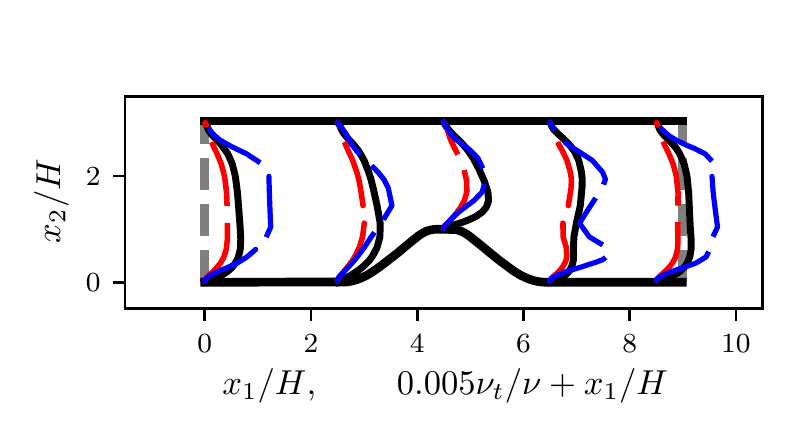}
            \caption{Profiles of eddy viscosity}
            \label{fig:hills_aposteriori:nut_prof}
        \end{subfigure}%
        \begin{subfigure}[b]{0.49\linewidth}
          \centering
            \includegraphics[trim=0.1in 0.15in 0.1in 0.3in,clip,scale=0.96]{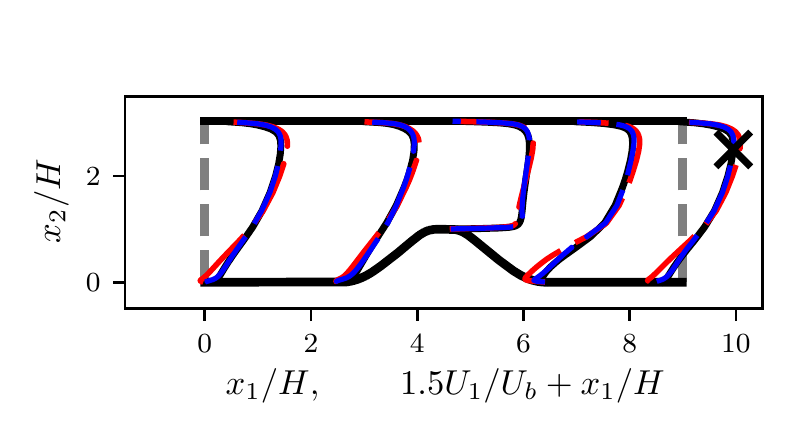}
            \caption{Profiles of streamwise velocity}
            \label{fig:hills_aposteriori:vel_prof}
        \end{subfigure}%
        \caption{Results of the RANS field inversion test case, reproduced with permission from Michelén~Ströfer et al.~\cite{michelenstrofer2019enforcing}. With a single observation of the velocity field the entire velocity field can be improved. While the inferred eddy viscosity field is closer to the truth than the prior it is still not correct due to the ill-posedness of the problem.}
        \label{fig:hills_aposteriori}
    \end{figure}

    The Reynolds averaged Navier--Stokes equations describe the mean velocity and pressure of fluid flows. 
    The RANS equations with eddy viscosity approximation of the Reynolds stress are given by 
    \begin{subequations}
            \label{eq:rans}
            \begin{gather}
                \frac{\partial U_i}{\partial x_i} = 0 \\
                U_j \frac{\partial U_i}{\partial x_j} = - \frac{\partial p}{\partial x_i} + \frac{\partial }{\partial x_i} \left((\nu+\nu_\text{t}) \left(\frac{\partial U_i}{\partial x_j} + \frac{\partial U_j}{\partial x_i} \right) \right) ,
            \end{gather}
        \end{subequations}
    where $U_i$ are the velocity components, $p$ is the pressure, $x_i$ are the spatial coordinates, $\nu$ is the fluid's kinematic viscosity, and $\nu_{\text{t}}$ is the unclosed eddy viscosity field. 
    This subsection presents the results, reproduced with permission from Michelén~Ströfer et al.~\cite{michelenstrofer2019enforcing}, of inferring the eddy viscosity field from a single point observation of the velocity for the two-dimensional flow over periodic hills. 
    The hill geometry is the same presented earlier in Figure~\ref{fig:phills}, the Reynolds number, based on hill height and bulk velocity, is $5,600$, and the mesh consists of $3,000$ cells.  
    The mode decomposition and reduced order modeling described earlier is used and the state to be inferred consists of the coefficients for the first $192$ modes. 
    The prior and synthetic truth solutions are obtained from solving the problem with different turbulence models. 
    For full details refer to Michelén~Ströfer et al.~\cite{michelenstrofer2019enforcing}. 
    The results using the EnKF are shown in Figure~\ref{fig:hills_aposteriori}. 
    It can be seen that a velocity observation at a single point results in a posterior velocity field that is much closer to the truth than the prior. 
    The inferred eddy viscosity field, while closer in magnitude to the truth, still deviates significantly from the ground truth. 
    This is due to the ill-posedness of the problem and could be addressed with further regularization. 
    
    \subsection{Uncertainty Quantification}
    \label{sec:cases:uq}
        One advantage of ensemble-based Bayesian methods is that they result in an estimate of the full posterior distribution. 
        This is in contrast to derivative based Bayesian methods where only the maximum a-posteriori estimate of the posterior is obtained. 
        However, for field inversion problems using the iterative EnKF, repeated applications of the EnKF tends to make samples collapse.
        This results in the proper mean value but misrepresents the true uncertainty in the estimate. 
        Zhang et al.~\cite{zhang2020uq} showed the better capabilities of EnKF-MDA and EnRML over EnKF to capture the uncertainty of the posterior. 
        They compared the results of the three methods to more accurate estimate of the posterior using a Markov chain Monte Carlo (MCMC) approach.  

        This test case demonstrate the use of DAFI for uncertainty quantification and highlights the advantage of methods like EnKF-MDA and EnRML over EnKF. 
        The test problem is a simple inversion problem consisting of two scalar states $\vect{x}=[\vecti{x}_1,\vecti{x}_2]^\top$, and two observations related to the the state by the observation operator 
        \begin{equation}
            \mapnl{H}(\vect{x}) = \begin{bmatrix} \vecti{x}_1 \\ \vecti{x}_1 + \vecti{x}_2^3 \end{bmatrix}
            \label{eq:uqobs}
        \end{equation}
        The prior distribution is given as 
        \begin{equation}
            \vect{x} \sim \mathcal{N}\left([0.5, 0.5]^\top, \operatorname{diag(0.1^2, 0.1^2)} \right) \text{.}
            \label{eq:uqprior}
        \end{equation}
        There is a single observation with value $y=[0.8, 2.0]^\top$ and independent variance of $0.05^2$ for both observed quantities. 
        
        \begin{figure}[!htb]
            \begin{subfigure}{1.0\linewidth}
                \centering
                \includegraphics[trim=0 130 0 0, clip]{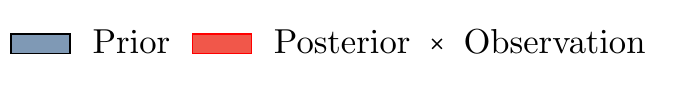}
            \end{subfigure}

        	\begin{subfigure}{0.33\linewidth}
        		\centering
        		\includegraphics[scale=0.9]{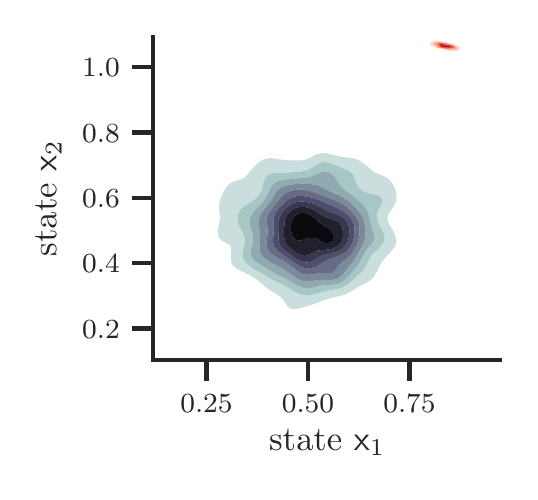}
                \caption{Iterative EnKF: State Space}
                \label{fig:uq:enkf_ss}
        	\end{subfigure}%
        	\begin{subfigure}{0.33\linewidth}
        		\centering
                \includegraphics[scale=0.9]{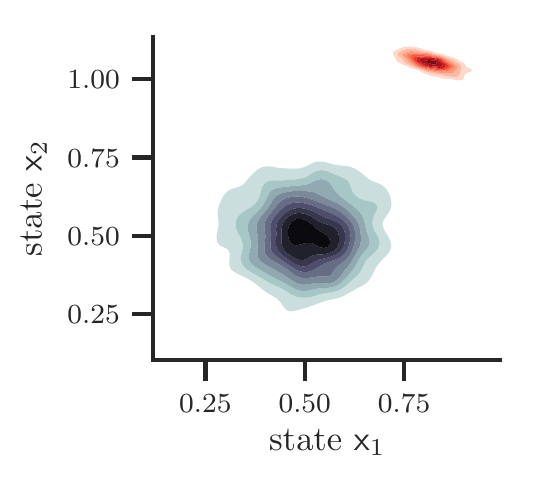}
                \caption{EnKF-MDA: State Space}
                \label{fig:uq:mda_ss}
        	\end{subfigure}%
        	\begin{subfigure}{0.33\linewidth}
        		\centering
                \includegraphics[scale=0.9]{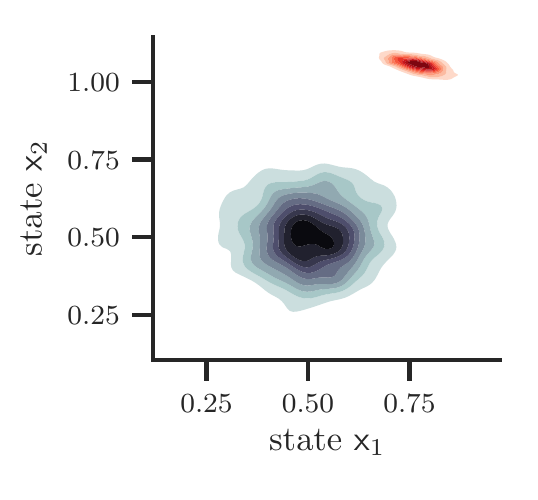}
                \caption{EnRML: State Space}
                \label{fig:uq:rml_ss}
        	\end{subfigure} 
        	
        	\begin{subfigure}{0.33\linewidth}
        		\centering
            \includegraphics[scale=0.9]{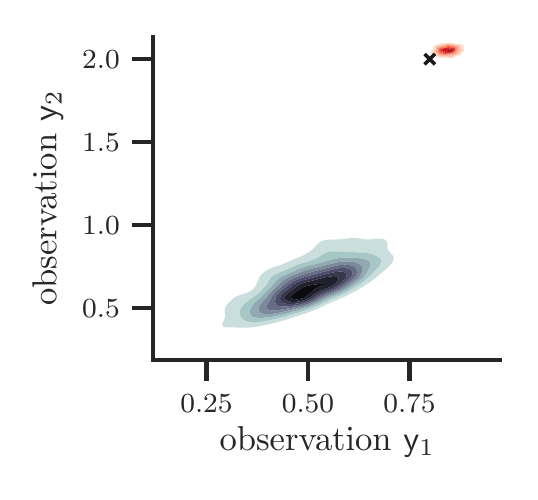}
                \caption{It. EnKF: Obs. Space}
                \label{fig:uq:enkf_obs}
        	\end{subfigure}%
        	\begin{subfigure}{0.33\linewidth}
        		\centering  
                \includegraphics[scale=0.9]{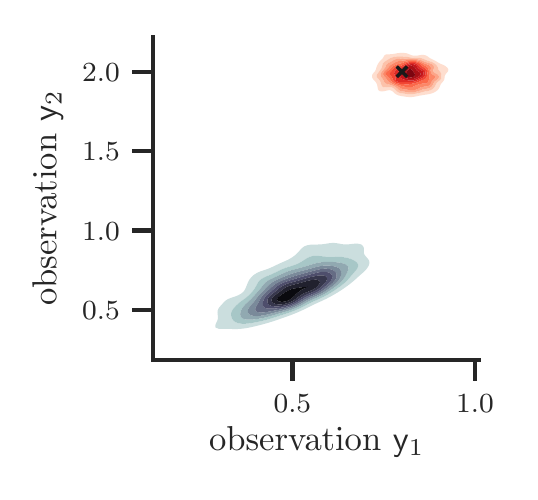}
                \caption{EnKF-MDA: Obs. Space}
                \label{fig:uq:mda_obs}
        	\end{subfigure}%
        	\begin{subfigure}{0.33\linewidth}
        		\centering
                \includegraphics[scale=0.9]{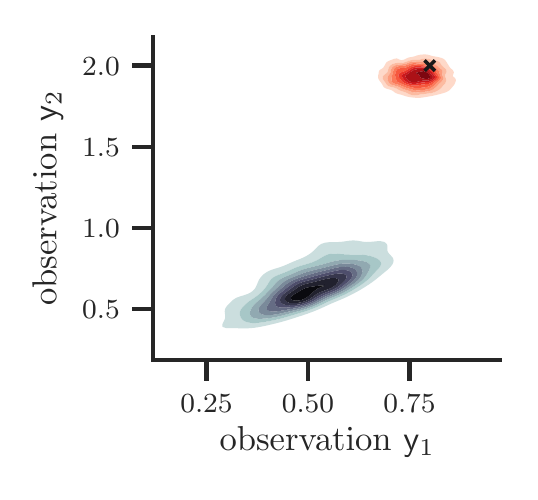}
                \caption{EnRML: Obs. Space}
                \label{fig:uq:rml_obs}
        	\end{subfigure}
        	\caption{Results of the uncertainty quantification test case. The results are shown in both the state space (upper row) and observation space (lower row) for all three methods. It is clear that the EnKF collapses, underestimating the uncertainty in the posterior distribution. The other two methods, and the EnKF-MDA especially, can better capture this uncertainty.}
        	\label{fig:uq}
        \end{figure}
        
        \begin{figure}[!htb]
            \begin{subfigure}{0.49\linewidth}
        		\centering
        		\includegraphics[scale=0.95]{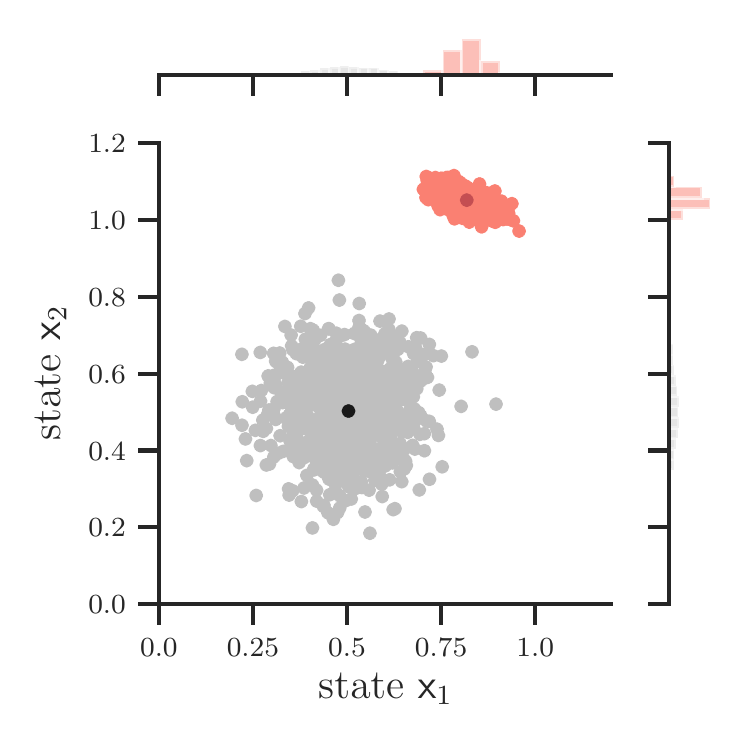}
                \caption{Samples}
                \label{fig:uq_mda:samps}
        	\end{subfigure}%
        	\begin{subfigure}{0.49\linewidth}
        		\centering
        		\includegraphics[scale=0.95]{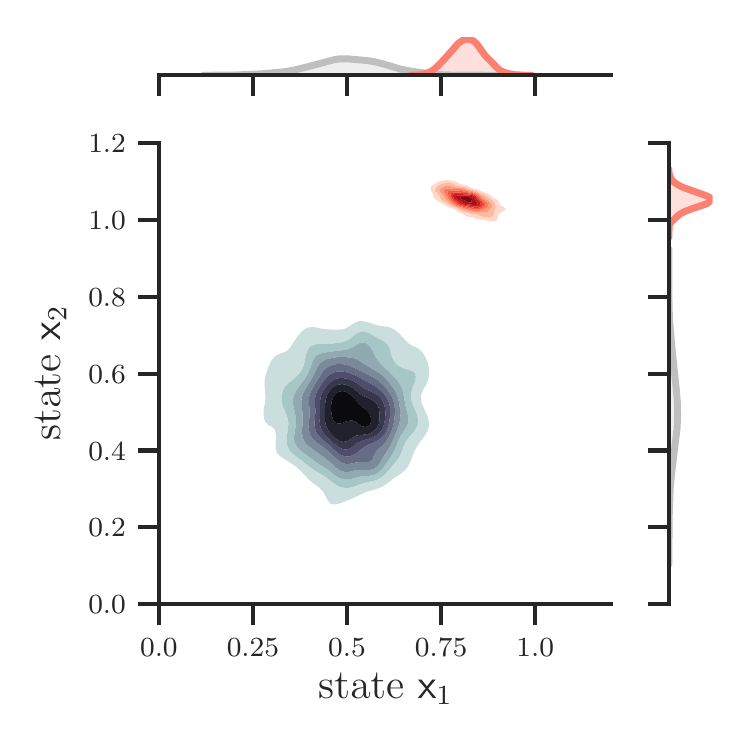}
                \caption{KDE}
                \label{fig:uq_mda:kde}
        	\end{subfigure}%
            \caption{Results for the uncertainty quantification test case using the EnKF-MDA. The results are shown in state space. The prior and posterior distributions are shown as well as their marginal distributions. Panel (a) shows the samples and sample mean and Panel (b) shows the fitted kernel density estimate (KDE).}
            \label{fig:uq_mda}
        \end{figure}
        
        The case was solved with the iterative EnKF, EnKF-MDA, and EnRML using $1000$ samples. 
        The iterative EnKF and the EnRML used the discrepancy principle for the stopping criteria with parameter $\tau=1.2$, and the EnRML used a length scale parameter of $\gamma=0.5$. 
        The iterative EnKF and EnRML converged after $7$ and $6$ iterations, respectively. 
        The EnKF-MDA was done with an inflation parameter of $N_{mda}=10$. 
        It was verified that larger values of the inflation parameter resulted in similar posterior distributions.
        The results are shown in Figure~\ref{fig:uq} where the kernel density estimate (KDE) for samples is plotted using the Scott method for the bandwidths. 
        All methods are successful in inferring mean closer to the observations. 
        The EnKF results do indeed collapse, which results in a poor estimate of the uncertainty. 
        Both the EnKF-MDA and EnRML are able to better capture the uncertainty. 
        Zhang et al.~\cite{zhang2020uq} demonstrate that while the shape of the posterior for the EnKF-MDA and EnRML might not be accurate, the estimated level of uncertainty is comparable to that obtained from MCMC. 
        The results for the EnKF-MDA in state space are shown again in Figure~\ref{fig:uq_mda} both as the raw samples and the fitted KDE. 
        This is done to elucidate the KDE and show more details on the marginal distributions.

\section{Software Installation and Numerical Examples}
\label{sec:install}
    The software can be obtained from the GitHub repository~\cite{dafi_github} or installed from the PyPI repository~\cite{dafi_pypi} using the Python package manager \texttt{pip}. 
    The code can run on any system with Python and NumPy and no compilation is required. 
    The computational requirements and running times depend greatly on the problem being solved which is user-specific. 
    The solution technique in this code requires running an ensemble of models and it is desirable to parallelize these evaluations. 
    The parallelization task falls under the user-defined physics model in either the \texttt{forecast\_to\_time} or \texttt{state\_to\_observation} functions, depending on the problem. 
    Through these functions the DAFI code provides the user's physics model an ensemble of states and expects back an updated ensemble. 
    The RANS field inversion tutorial for the periodic hill (Section~\ref{sec:cases:hills}) is also an example of how to parallelize this task using the standard Python library. 
    This example is able to run in HPC machines but as currently implemented it is limited to multiple cores in a single node. 
    More complex parallelization, such as paralleizing each of the individual model evaluations, are possible but fall under the user's physics model. 
    
    Several numerical examples were presented in Section~\ref{sec:cases}. 
    The complete setup including input files are included in the \texttt{tutorials} directory of the software distribution. 
    Detailed explanation of the input file are described in the documentation~\cite{dafi_rtd}. 
    In addition to the input file, the user needs to provide a physics model with predefined methods as detailed in Section~\ref{sec:implementation:physics}. 
    These files are included in the tutorials for all test cases presented and Appendix~\ref{app:code} provides some code example for such a file.

\section{Concluding Remarks}
\label{sec:conclusion}
    DAFI is an open-source, object-oriented Python package aimed at providing a robust framework for solving data assimilation and field inversion problems.
    It contains a library of ensemble-based, derivative-free, Bayesian methods based on the ensemble Kalman filter, and a straightforward way of coupling the user's domain-specific model.
    As such the user is required to have domain-specific knowledge but little knowledge is required on the data assimilation and field inversion techniques.
    As an example, the authors have used DAFI to infer the Reynolds stress tensor in Reynolds-averaged Navier--Stokes (RANS) simulations of fluid flow\cite{xiao_quantifying_2016}. 
    Alternatively DAFI can be used as a test-bed for research on new inversion methods, which is straightforward due to the object-oriented nature of the code.
    As examples, the authors have used DAFI to develop the regularized ensemble Kalman filter~\cite{zhang2019regularization} and to enforce boundary conditions during field inversion~\cite{michelenstrofer2019enforcing}.
    One distinguishing feature of DAFI is that it was created with physical fields in mind.
    As such it includes several modules for field operations such as creating covariance matrices using different kernels, or performing the Karhunen-Loève decomposition. 
    All this can be done in unstructured discretization of the fields. 

    This paper presents the theory behind ensemble-based methods and random fields, the approach taken to generalize diverse problems into a general problem, and the code implementation.
    It also presents several test cases that show the use of DAFI as well as the diversity in the types of problems it is meant to solve.
    The test cases include a scalar data assimilation problem using the Lorenz equations, a regularized field inversion problem using the diffusion equation, and an uncertainty quantification problem. 
    The results from a different study, using DAFI, for a practical field inversion problem using the RANS equations are also shown, with permission.
    The code is actively maintained in GitHub~\cite{dafi_github}, released through PyPI~\cite{dafi_pypi}, and the documentation is available online through Read the Docs~\cite{dafi_rtd}.
    The GitHub repository and the documentation include several tutorials.
    The authors hope the code is beneficial to other researchers and in keeping with the open source nature of the project accept code improvements through GitHub pull requests.

\section*{Acknowledgements}
    Part of DAFI was adopted from previous code developed at Virginia Tech by Dr. Jianxun Wang, Dr. Jinlong Wu, and Dr. Rui Sun, whose contributions are gratefully acknowledged. 
    The authors would like to thank Dr. Gangfeng Ma at Old Dominion University for his valuable feedback as one of the earliest external adopters of the DAFI code.
    The authors would also like to thank Kristopher Olshefski for his help reviewing this manuscript.

\appendix

\section{Kalman Gain Matrix for EnKF}
\label{app:kalmangain}
    This appendix shows how the Kalman gain matrix for the EnKF can be written in terms of the state mapped to observation space $\vect{z}=\mapl{H}\vect{x}$ by using the definition of sample covariance.
    This avoids using, or having to explicitly construct, the observation operator $\mapl{H}$.
    It will also avoid having to construct, or perform matrix multiplications with, the full state covariance matrix $C_{\vect{x}}$.
    The state $\vect{x}$ mapped to observation space is
    \begin{equation}
        \vect{z} = \mapl{H}\vect{x} \text{.}
    \end{equation}
    However the vector $\vect{z}$ can typically be obtained directly from the physics model without explicitly constructing the matrix $\mapl{H}$ or carrying out the matrix multiplication.
    The Kalman gain matrix for the EnKF is given by
    \begin{equation}
        \mapl{K} = C_{\vect{x}}\mapl{H}^\top\left(\mapl{H}C_{\vect{x}}\mapl{H}^\top + C_{\vect{y}} \right)^{-1} \text{,}
        \label{eq:app:k}
    \end{equation}
    where the sample covariance for the state $\vect{x}$ is given by
    \begin{equation}
        C_{\vect{x}} = \frac{1}{N_s-1} \sum_{j=1}^{N_s} (\vect{x}^{(j)}-\overline{\vect{x}})(\vect{x}^{(j)}-\overline{\vect{x}})^\top \text{,}
        \label{eq:app:covx}
    \end{equation}
    and $N_s$ is the number of samples in the ensemble, $\vect{x}^{(j)}$ is a particular sample, and $\overline{\vect{x}}$ is the sample mean.
    Substituting Equation\eqref{eq:app:covx} into Equation\eqref{eq:app:k} and rearranging, the Kalman gain matrix can be written as
    \begin{equation}
        K = C_{\vect{x}\vect{z}} \left(C_{\vect{z}} + C_{\vect{y}} \right)^{-1} \text{,}
        \label{eq:app:k2}
    \end{equation}
    where $C_{\vect{y}}$ is the observation error (covariance) matrix, $C_{\vect{z}}$ is defined similarly to $C_{\vect{x}}$, and the covariance matrix between the state and the state mapped to observation space is given by
    \begin{equation}
        C_{\vect{x}\vect{z}} = \frac{1}{N_s-1} \sum_{j=1}^{N_s} (\vect{x}^{(j)}-\overline{\vect{x}})(\vect{z}^{(j)}-\overline{\vect{z}})^\top \text{.}
        \label{eq:app:covxz}
    \end{equation}
    Note that the Kalman gain in Equation~\eqref{eq:app:k2} does not include the very large state covariance matrix $C_{\vect{x}}$, rather it involves much smaller covariance matrices since the dimension of the observation space is generally much smaller than that of the state space.

\section{Iterative EnKF for Field Inversion}
\label{app:iterativekalman}
    In field inversion problems one field is related to another through a nonlinear forward operator, and the input field is inferred from observations of the output field.
    This appendix shows how a field inversion problem can be recast as an artificial dynamics problem using an augmented state and the EnKF used to solve it~\cite{iglesias_ensemble_2013}.
    The augmented state vector written in terms of block matrices is
    \begin{equation}
        \widetilde{\vect{x}} = \begin{bmatrix} \vect{x}\\ \mapnl{H}(\vect{x}) \end{bmatrix} = \begin{bmatrix} \vect{x}\\ \vect{z} \end{bmatrix} \text{,}
    \end{equation}
    and \emph{linear} observation operator is
    \begin{equation}
        \widetilde{\mapl{H}}=\begin{bmatrix}\bm{0} & \mathrm{I}\end{bmatrix} \text{.}
    \end{equation}
    Here the state $\vect{x}$ mapped to observation space is $\vect{z}$ and is given by a nonlinear observation operator $\mapnl{H}$ that combines the forward model and the observation operator.
    The artificial dynamics is given by
    \begin{equation}
        \widetilde{\vect{x}}_{i+1} = \widetilde{\mapnl{M}}(\widetilde{\vect{x}}_i) = \begin{bmatrix} \vect{x}_i \\ \mapnl{H}_i(\vect{x}_i) \end{bmatrix} \text{.}
    \end{equation}
    The analysis state is then given by
    \begin{equation}
        \widetilde{\vect{x}}^{a(j)} = \widetilde{\vect{x}}^{f(j)} + \widetilde{K}\left(\vect{y}^{(j)}-\widetilde{\mapl{H}}\widetilde{\vect{x}}^{f(j)}\right)  \text{,}
        \label{eq:appb:enkf}
    \end{equation}
    with Kalman gain matrix
    \begin{equation}
        \widetilde{\mapl{K}} = C_{\widetilde{\vect{x}}}\widetilde{\mapl{H}}^\top\left(\widetilde{\mapl{H}}C_{\widetilde{\vect{x}}}\widetilde{\mapl{H}}^\top + C_{\vect{y}} \right)^{-1} \text{,}
        \label{eq:appb:k}
    \end{equation}
    where the the covariance matrix is composed of four blocks as
    \begin{equation}
      C_{\widetilde{\vect{x}}} = \begin{bmatrix} C_{\vect{x}} & C_{\vect{x}\vect{z}} \\ C_{\vect{x}\vect{z}}^\top & C_{\vect{z}} \end{bmatrix} \text{.}
      \label{eq:appb:c}
    \end{equation}

    The analysis step can be expressed in terms of the original state $\vect{x}$.
    Substituting Equation~\eqref{eq:appb:c} into Equation~\eqref{eq:appb:k} and carrying out the block operations the Kalman gain matrix becomes
    \begin{equation}
        \widetilde{K} = \begin{bmatrix} C_{\vect{x}\vect{z}}(C_{\vect{z}}+C_{\vect{y}})^{-1} \\ C_{\vect{z}}(C_{\vect{z}}+C_{\vect{y}})^{-1} \end{bmatrix} \text{,}
    \end{equation}
    and Equation~\eqref{eq:appb:enkf} becomes
    \begin{equation}
        \begin{bmatrix} \vect{x}^{a(j)}\\ \vect{z}^{a(j)} \end{bmatrix} = \begin{bmatrix} \vect{x}^{f(j)}\\ \vect{z}^{f(j)} \end{bmatrix} + \begin{bmatrix} C_{\vect{x}\vect{z}}(C_{\vect{z}}+C_{\vect{y}})^{-1} \\ C_{\vect{z}}(C_{\vect{z}}+C_{\vect{y}})^{-1} \end{bmatrix} \begin{bmatrix} \vect{y}^{(j)} - \mapnl{H}(\vect{x}^{f(j)}) \end{bmatrix}
    \end{equation}
    or
    \begin{subequations}
        \begin{gather}
            \vect{x}^{a(j)} = \vect{x}^{f(j)} + C_{\vect{x}\vect{z}}(C_{\vect{z}}+C_{\vect{y}})^{-1} \left(\vect{y}^{(j)} - \mapnl{H}(\vect{x}^{f(j)})\right) \text{,} \label{eq:appb:enkf2}\\
            \vect{z}^{a(j)} = \vect{z}^{f(j)} + C_{\vect{z}}(C_{\vect{z}}+C_{\vect{y}})^{-1} \left(\vect{y}^{(j)} - \mapnl{H}(\vect{x}^{f(j)})\right) \text{.} \label{eq:appb:enkf2b}
        \end{gather}
    \end{subequations}
    Equation~\eqref{eq:appb:enkf2} is the EnKF update scheme but with nonlinear observation operator:
    \begin{equation}
        \vect{x}^{a(j)} = \vect{x}^{f(j)} + K \left(\vect{y}^{(j)} - \mapnl{H}(\vect{x}^{f(j)})\right) \text{.}
    \end{equation}
    That is, the artificial dynamics can be solved using EnKF on the original state vector but using the nonlinear observation operator and iterating by using the updated analysis state as the forecast state until convergence is achieved.
    Note that Equation~\eqref{eq:appb:enkf2b} is not needed and solving the full augmented system would include unnecessary computations.

\section{Discrete KL Modes}
\label{app:kl}
    When the state vector includes physical fields these need to be discretized, using some mesh which is generally unstructured. 
    This appendix describes how to obtain the KL modes in the discrete, unstructured mesh, case. 
    The continuous Fredholm integral equation 
     \begin{equation}
        \int_\Omega \mathcal{C}(\bm{\xi_1},\bm{\xi_2})e_k(\bm{\xi_1})\mathrm{d}\bm{\xi_1} = \lambda_k e_k(\bm{\xi_2}) \text{}
        \label{eq:klint}
    \end{equation}
    can be discretized as 
    \begin{equation}
        \sum_{i=1}^N C_{ij}e_k\Delta\!\Omega_i = \lambda_k e_{k,j} \text{,}
    \end{equation}
    where $N$ is the number of cells in the discretization, $\Delta\!\Omega_i$ is the volume of the $i$\textsuperscript{th} cell and $C$ and $e_k$ are now discrete covariance matrix and eigevector.
    These $N$ equations (for $j\in[1,N]$) can be written as a matrix equation as
    \begin{equation}
        \left( CW \right) e_k = \lambda_k e_k \label{eq:kldisc1} \text{,} \\
    \end{equation}
    with 
    \begin{equation}
        W = \begin{bmatrix} 
                \Delta\!\Omega_1 &  &  \\ 
                 & \ddots & \\
                 & & \Delta\!\Omega_N
            \end{bmatrix} \text{,}
    \end{equation}
    which is an eigenvalue problem. 
    The discrete KL modes are then given by 
    \begin{equation}
        \phi_k = \sqrt{\lambda_k}\hat{e}_k \text{,}
    \end{equation}
    where $\hat{e}_k$ are the normalized eigenvectors (using the norm in Equation~\eqref{eq:discretenorm}).

    Optionally, the basis can be changed such that the standard vector dot product in the new basis is equivalent to the weighted L2 norm of Equation~\eqref{eq:discretenorm} in the original basis.
    The desired change of basis matrix $L$, which transforms a vector from the original coordinates to the desired coordinates, is such that  
    \begin{equation}
        x^\top W x = (Lx)^\top (Lx) \text{.}
    \end{equation}
    This can be written as 
    \begin{equation}
        x^\top W x = x^\top L^\top L x \text{,}
    \end{equation}
    which leads to 
    \begin{gather}
        W = L^\top L\text{,} \\
        L = L^\top = \begin{bmatrix} 
                        \sqrt{\Delta\!\Omega_1} &  &  \\ 
                        & \ddots & \\
                        & & \sqrt{\Delta\!\Omega_N}
                    \end{bmatrix} \text{,} \\ 
        W = L L
    \end{gather}
    Changing the basis of both sides of Equation~\eqref{eq:kldisc1} results in 
    \begin{gather}
        L \left( CW \right) e_k = L \lambda_k e_k \text{,} \\
        L C L L e_k = \lambda_k L e_k \text{,} \\
        \left(L C L\right) \left(L e_k\right) = \lambda_k \left(L e_k\right) \label{eq:kldisc2} \text{.}
    \end{gather}
    This is still the same eigenvalue problem but in different coordinates. 
    The eigendecomposition of $LCL$ gives the eigenvalues and the eigenvectors $g_k=L e_k$ which can be normalized using the standard dot product to $\hat{g}_k$. 
    The normalized (weighted norm definition) eigenvectors are then obtained by reverting to the original basis as 
    \begin{equation}
        \hat{e}_k = (L)^{-1}\hat{g}_k \text{.}
    \end{equation}

\section{Code Example}
\label{app:code}
    This appendix presents a simple example of a physics model. 
    The physics model is that used for the scalar inversion problem in Section~\ref{sec:cases:uq}. 
    In particular, note the implementation of the required method described in Section~\ref{sec:implementation:physics}. 
    The \texttt{forecast\_to\_time} method is not implemented since this field inversion problem does not requires it.
    The model is first initialized via the \texttt{\_\_init\_\_} method using the following input dictionaries. 
    \begin{python}
inputs_dafi = {
    'model_file': 'model.py',
    'inverse_method': 'EnKF',
    'nsamples': 1000,
    'max_iterations': 100,
    'convergence_option': 'discrepancy',
    'convergence_factor': 1.2,
}

inputs_model = {
    'x_init_mean': [0.5, 0.5],
    'x_init_std': [0.1, 0.1],
    'obs': [0.8, 2.0],
    'obs_std': [0.05, 0.05],
}
    \end{python}
    The DAFI input defines the number of samples, physics model and inverse method to use, and convergence criteria. 
    The model input defines the prior distribution and the observation data. 
    The initial ensemble is generated in the \texttt{generate\_ensemble} method using the specified prior mean and standard deviation as in Equation~\eqref{eq:uqprior}. 
    The \texttt{state\_to\_observation} method maps the ensemble from state space to observation space based on the observation operator in Equation~\eqref{eq:uqobs}. 
    Finally, the \texttt{get\_obs} method simply returns the value of the observation and the observation error. 
    The complete physics model is shown below. 
    The physics model and input files for all test cases presented here are included in the tutorials distributed with the code. 
    
    \begin{python}
# Copyright 2020 Virginia Polytechnic Institute and State University.
""" Dynamic model for solving the scalar inversion problem used for
uncertainty quantification. """

# standard library imports
import os

# third party imports
import numpy as np
import yaml

# local imports
from dafi import PhysicsModel

class Model(PhysicsModel):

    def __init__(self, inputs_dafi, inputs_model):
        # save the required inputs
        self.nsamples = inputs_dafi['nsamples']

        # read inputs
        self.init_state = np.array(inputs_model['x_init_mean'])
        self.state_std =  np.array(inputs_model['x_init_std'])
        self.obs =  np.array(inputs_model['obs'])
        self.obs_err = np.diag(np.array(inputs_model['obs_std'])**2)

        # required attributes.
        self.name = 'Scalar Inversion Case for UQ'

        # other attributes
        self.nstate = len(self.init_state)

    def __str__(self):
        return self.name

    def generate_ensemble(self):
        state_vec = np.empty([self.nstate, self.nsamples])
        for i in range(self.nstate):
            state_vec[i,:] = np.random.normal(
                self.init_state[i], self.state_std[i], self.nsamples)
        return state_vec

    def state_to_observation(self, state_vec):
        obs_1 = state_vec[0, :]
        obs_2 = state_vec[0, :] + state_vec[1, :]**3
        model_obs = np.array([obs_1, obs_2])
        return model_obs

    def get_obs(self, time):
        return self.obs, self.obs_err
    \end{python}

\bibliographystyle{elsarticle-num}
\bibliography{references}

\end{document}